\documentclass[12pt]{article}
\input{preamble}
\newcommand{\blind}{0}

\begin{document}

\def\spacingset#1{\renewcommand{\baselinestretch}%
{#1}\small\normalsize} \spacingset{1}

\if0\blind
{
  \title{\bf Proximal MCMC for Bayesian Inference
of Constrained and Regularized Estimation}
  \author{Xinkai Zhou \hspace{.2cm}\\
    \small{Department of Biostatistics, UCLA}\\
    Qiang Heng \hspace{.2cm}\\
    \small{Department of Computational Medicine, UCLA}\\
    Eric C. Chi \hspace{.2cm}\\
    \small{Department of Statistics, Rice University}\\
    and \\
    Hua Zhou \hspace{.2cm}\\
    \small{Department of Biostatistics, UCLA}\\}
    \date{}
  \maketitle
} \fi

\if1\blind
{
  \bigskip
  \bigskip
  \bigskip
  \begin{center}
    {\LARGE\bf Proximal MCMC for Bayesian Inference
of Constrained and Regularized Estimation}
\end{center}
  \medskip
} \fi

\bigskip

\begin{abstract}
This paper advocates proximal Markov Chain Monte Carlo (ProxMCMC) as a flexible and general Bayesian inference framework for constrained or regularized estimation. Originally introduced in the Bayesian imaging literature, ProxMCMC employs the Moreau-Yosida envelope for a smooth approximation of the total-variation regularization term, fixes variance and regularization strength parameters as constants, and uses the Langevin algorithm for the posterior sampling. We extend ProxMCMC to be fully Bayesian by providing data-adaptive estimation of all parameters including the regularization strength parameter. More powerful sampling algorithms such as Hamiltonian Monte Carlo are employed to scale ProxMCMC to high-dimensional problems. Analogous to the proximal algorithms in optimization, ProxMCMC offers a versatile and modularized procedure for conducting statistical inference on constrained and regularized problems. The power of ProxMCMC is illustrated on various statistical estimation and machine learning tasks, the inference of which is traditionally considered difficult from both frequentist and Bayesian perspectives.
\end{abstract}

\noindent%
\vfill

\newpage
\spacingset{1}

\section{Introduction}
Many statistical learning tasks are posed as regularized maximum likelihood estimation problems, which require solving optimization problems of the form
$$
\text{maximize} \quad \ell(\thetabf) - \rho P(\thetabf),
$$
where $\thetabf$ denotes model parameters, $\ell(\thetabf)$ denotes the log-likelihood and quantifies the lack-of-fit between the model and the data, $P(\thetabf)$ is a regularization function that imposes  structure on parameter estimates, and $\rho$ is a nonnegative regularization strength parameter that trades off the model fit encoded in $\ell(\thetabf)$ with the desired structure encoded in $P(\thetabf)$. Canonical examples of regularization functions include the $\ell_1$-norm that promotes sparsity and the nuclear norm that promotes recovery of low-rank models. To date, most work has focused exclusively on estimating $\thetabf$ without quantifying the uncertainty in the estimates. Lacking tools for assessing uncertainty in findings from regularized models, practitioners often resort to classical inference tools designed for non-regularized models. This practice will substantially inflate the type I error and lead to unreproducible scientific discoveries. 

This issue has motivated the development  of post-selection inference techniques such as simultaneous inference \citep{BerkBrownBujaZhangZhao12PostSelectionInference, BachocPreinerstorferSteinberger2020PostSelectionInference, Kuchibhotla20PostSelectionInference} and selective inference \citep{LeeSunSunTaylor2016PostSelectionInference, choi2017postpca, taylor2018postglm}. A closely related approach calculates confidence intervals for coefficients of high-dimensional linear models through bias-correction \citep{vandeGeerBuhlmannRitovdezeure14DebaisedLasso, ZhangZhang14DebiasedLasso, JavanmardMontanari14DebiasedLasso}.  
Most of this literature, however, focuses on variable selection through the $\ell_1$-regularization. Extending these strategies to other regularizations and to problems involving constraints is not straightforward. Moreover, caution is warranted when reporting these confidence intervals because their interpretation (e.g., conditional on the selection event) differs from traditional ones.

An alternative is to cast the problem in the Bayesian framework. For example, \citet{TrevorCasella08BayesianLasso} introduced the Bayesian lasso, where the $\ell_1$-regularization was identified with a Laplace prior and a Gibbs sampler was used to sample from the posterior distribution. This work is part of a large literature on Bayesian variable selection methods, which include sparsity inducing prior distributions such as spike-and-slab \citep{MitchellBeauchamp88BayesianVariableSelection,GeorgeMcCulloch93SpikeSlab}, horseshoe \citep{CarvalhoPolsonScott10Horseshoe, PolsonScott10ShrinkagePriors,piironen2017horseshoe, bhadra2019lasso}, orthant normal \citep{Hans11ShrinkagePrior}, correlated Normal-Gamma \citep{GriffinBrown12ShrinkagePrior, GriffinBrown13ShrinkagePrior}, generalized double Pareto \citep{Armagan11ShrinkagePrior}, and Dirichlet-Laplace \citep{BhattacharyaPatiPillaiDunson15ShrinkagePrior}. Despite constant innovations in Bayesian techniques for variable selection, incorporating regularizations and constraints beyond sparsity still requires a substantial amount of problem-specific analysis.

More recently, \citet{pereyra2016proximal} and \citet{durmus2018efficient,DurmusMoulinesPereyra22ProxMCMC} proposed the proximal Markov Chain Monte Carlo (ProxMCMC) algorithm for quantifying uncertainty in Bayesian imaging applications where the regularizations of interest include the total-variation semi-norm \citep{RudinOsherFatemi1992} and the $\ell_1$-norm. To deal with the non-smoothness of these regularizations, they employ the Moreau-Yosida envelope to obtain their smooth approximations. Samples from the smooth approximate posterior distribution can be drawn using Langevin dynamics. Their approach offers a framework for conducting statistical inference on regularized regression models whenever the regularization term is convex and admits a proximal map that can be computed efficiently, which holds true for a wide variety of regularizations. The fly in the ointment, however, is that their approach requires manually setting the regularization strength parameter $\rho$. One solution to this problem is given by \citet{vidal2020maximum} and \citet{de2020maximum}, who proposed using an empirical Bayes method called the stochastic approximation proximal gradient (SAPG) to estimate the regularization strength parameter by maximum marginal likelihood. It only provides point estimates of the regularization strength parameter, potentially resulting in suboptimal statistical precision due to the neglect of uncertainty in the regularization strength parameter. In terms of flexibility, the SAPG approach focuses on regularized estimation problems, while constrained estimation problems remain relatively under-explored.  

In this paper, we address this limitation and extend ProxMCMC to be fully Bayesian by incorporating regularization and constraints through epigraph priors. Our extended ProxMCMC inference framework is suitable for regularized or constrained statistical learning problems and offers three main advantages. First, it provides valid and automatic statistical inference even for problems that involve non-smooth and potentially non-convex regularization or constraints. The inference for such problems is traditionally considered difficult. Second, it is fully Bayesian, eliminating the need for parameter tuning. This is in contrast to previous ProxMCMC methods \citep{durmus2018efficient,DurmusMoulinesPereyra22ProxMCMC} where the regularization strength parameter is either manually fixed or requires tuning. Third, the method is highly modular. Its components -- model, prior, proximal map, and sampling algorithm -- are independent of each other and can be easily adjusted to address new problems. This feature makes ProxMCMC highly customizable, allowing users to tailor it to their specific problems. 
The practical significance of the last point cannot be emphasized enough and is exemplified in the constrained lasso example, where the ``sum to zero" constraint, imposed by problem-specific considerations, causes existing inference methods to break down, but poses no challenge for the proposed ProxMCMC method. We will save the details for Section \ref{sec:constrained-lasso}.

Finally, we put the proposed ProxMCMC method on firm foundations by providing guarantees on the properness of the approximate posterior and showing that the approximate posterior can be made arbitrarily close to the target posterior in total-variation under suitable assumptions.

The rest of the paper is organized as follows. Section 2 reviews concepts from convex optimization that form the building blocks of the ProxMCMC framework. Section 3 illustrates our method using the familiar lasso problem. Section 4 summarizes the key elements from our case study of lasso to show how the ProxMCMC method can be applied generally. Section 5 presents a variety of illustrative applications, whose numerical results are presented in Section 6.  Sections 7 provides a brief discussion, while theoretical guarantees can be found in the supplementary materials.

\section{Background}\label{sec:background}

We review concepts from convex analysis essential for ProxMCMC, specifically Moreau-Yosida envelopes and  proximal mappings. For a more thorough review of proximal mappings and their applications in statistics and machine learning, we refer readers to \citet{Combettes2005, Combettes2011, PolsonScott2015Proximal}. In convex optimization it is often convenient to work with functions that map into the extended reals, $\mathbb{\bar{R}} = \Real \cup \{\infty\}$. The indicator function of a set $C$, denoted $\delta_C(\xbf)$, is defined as 
\begin{eqnarray}\label{eqn:indicator-function}
\delta_C(\xbf) & = & \begin{cases}
0 & \xbf \in C \\
\infty & \text{otherwise,}
\end{cases}
\end{eqnarray}
which differs from the familiar $0/1$ indicator function used in statistics. A function $f: \mathbb{E} \rightarrow \mathbb{\bar{R}}$ is \emph{lower-semicontinuous} at $\xbf \in \mathbb{E}$ if 
\begin{eqnarray}\label{eqn:lower-semicontinuity}
f(\xbf) \leq \liminf_{n\rightarrow\infty} f(\xbf_n)
\end{eqnarray}
for any sequence $\{\xbf_n\}_{n\geq 1} \subseteq \mathbb{E}$ for which $\xbf_n \rightarrow \xbf$ as $n\rightarrow \infty$. A function is \emph{proper} if it takes on a finite value for some element in its domain. When the set $C$ is closed and convex, the indicator function $\delta_C(\xbf)$ is lower-semicontinuous and convex. Let $\Gamma(\Real^m)$ denote the set of all proper, lower-semicontinuous, convex functions from $\Real^m$ into $\mathbb{\bar{R}}$. The Euclidean norm of a point $\xbf$ is denoted using the familiar notation $\lVert \xbf \rVert$.

\subsection{Moreau-Yosida Envelopes and Proximal Maps}

\begin{definition}\label{def:prox-operator}
Given $g \in \Gamma(\Real^m)$ and a positive scaling parameter $\lambda$, the \emph{proximal mapping} of $g$ is the operator given by
\begin{eqnarray*}
\prox_g^\lambda(\xbf) &= & \underset{\omegabf}{\arg\min}\; \left\{g(\omegabf) + \frac{1}{2\lambda}\|\omegabf-\xbf\|^2\right\}.
\end{eqnarray*}
\end{definition}
\begin{definition}\label{def:moreau-envelope}
Given $g \in \Gamma(\Real^m)$ and a positive scaling parameter $\lambda$, the \emph{Moreau-Yosida envelope} of $g$ is given by
\begin{eqnarray*}
g^\lambda(\xbf) & = & \inf_{\omegabf} \left\{g(\omegabf) + \frac{1}{2\lambda}\|\omegabf-\xbf\|^2\right\}.
\end{eqnarray*}
\end{definition}
The infimum is always attained at a unique point when $g \in \Gamma(\Real^m)$, and the minimizer defines the proximal mapping of $g$.

Intuitively, evaluating the proximal mapping of $g$ at $\xbf$ identifies a point $\omegabf$ that balances between minimizing $g$ and staying close to $\xbf$ in Euclidean distance. The extent to which $\omegabf$ minimizes $g$ is controlled by the positive scaling parameter $\lambda$: larger values of $\lambda$ pushes $\omegabf$ closer to the minimum, whereas smaller values keep $\omegabf$ closer to $\xbf$. From the definition, we can see that the Moreau-Yosida envelope is related to the proximal mapping through the equation $g^\lambda(\xbf) = g(\prox_g^\lambda(\xbf)) + \frac{1}{2\lambda}\|\prox_g^\lambda(\xbf)-\xbf\|^2$. 

\begin{figure}[!tbp]
\centering
\includegraphics[width=0.49\textwidth]{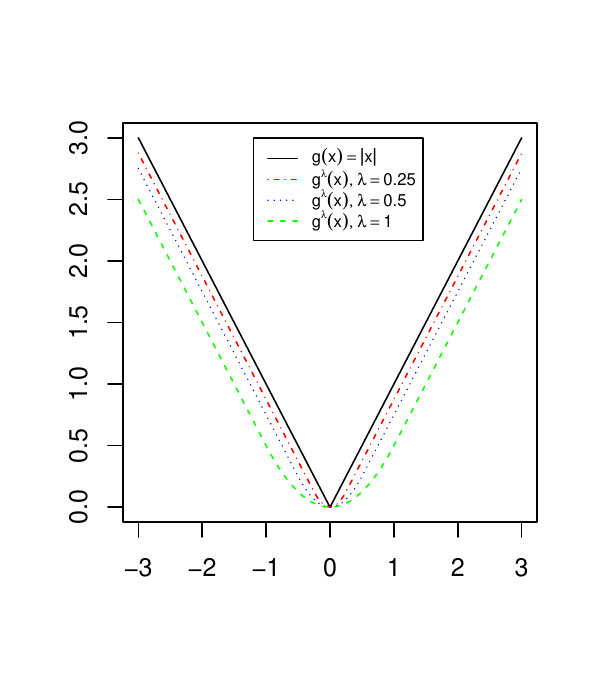}
\includegraphics[width=0.49\textwidth]{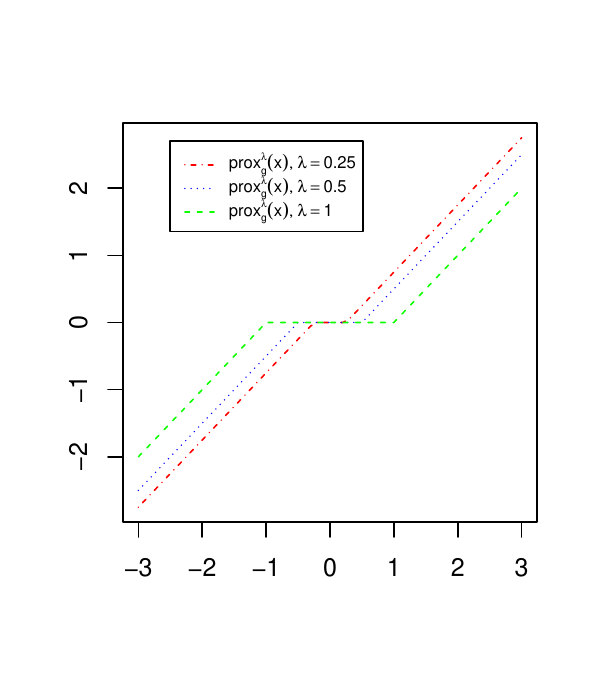}
  \caption{The Moreau-Yosida envelope (left) and proximal mapping (right) of the absolute value function $g(x) = \lvert x \rvert$.}
  \label{fig:prox-moreau}
\end{figure}

We illustrate these  definitions using the well known Huber function 
\begin{eqnarray*}
g^\lambda(x) & = & \begin{cases}
\frac{1}{2\lambda}x^2 & \text{if $\lvert x \rvert \leq \lambda$} \\
\lvert x \rvert - \frac{\lambda}{2} & \text{otherwise,}
\end{cases}
\end{eqnarray*}
which is the Moreau-Yosida envelope of the absolute value function $g(x) = \lvert x \rvert$. The left panel of Figure \ref{fig:prox-moreau} shows $g(x)$ and $g^\lambda(x)$ for three different $\lambda$ values. This familiar example from robust statistics shows that the Moreau-Yosida envelope provides a differentiable approximation to a non-smooth function where the approximation improves as $\lambda$ gets smaller. The corresponding proximal map is the celebrated soft-thresholding operator $S_\lambda(x)$ defined by
\begin{equation}
\label{eqn:soft-thresholding}
S_\lambda(x) = \begin{cases}
x - \lambda  & \text{if $x > \lambda$} \\
0 & \text{if $\lvert x \rvert \leq \lambda$} \\
x + \lambda & \text{if $x < -\lambda$.}
\end{cases}
\end{equation}
In the right panel of Figure \ref{fig:prox-moreau}, we show $\prox_g^\lambda(x)$ for the same $\lambda$ values as in the left panel.

In general, the Moreau-Yosida envelope $g^\lambda(\xbf)$ has several important properties. First, $g^\lambda(\xbf)$ is convex when $g(\xbf)$ is convex. Second, if $g(\xbf)$ is convex, then $g^\lambda(\xbf)$ is always differentiable even if $g(\xbf)$ is not, and its gradient can be expressed in terms of $\prox^\lambda_g(\xbf)$, namely,
\begin{eqnarray}
\label{eqn:moreau-gradient}
\nabla g^\lambda (\xbf) & = & \frac{1}{\lambda}\left[\xbf-\operatorname{prox}^\lambda_g(\xbf) \right].
\end{eqnarray}
Moreover, $\nabla g^\lambda(\xbf)$ is $\lambda^{-1}$-Lipschitz since proximal mappings are firmly nonexpansive. Finally, $g^\lambda(\xbf)$ converges pointwise to $g(\xbf)$ as $\lambda$ tends to zero~\citep{rockafellar2009variational}. In summary, the Moreau-Yosida envelope of a non-smooth function $g(\xbf)$ is a  Lipschitz-differentiable, arbitrarily close approximation to $g(\xbf)$.

The closely related proximal mapping plays a prominent role in modern statistical learning since many popular non-smooth regularizations have unique proximal maps that either have explicit formulas or can be computed efficiently \citep{Beck17FOM}.  

In the special case when $g$ is the indicator function $\delta_\mathcal{E}$ of a set $\mathcal{E}$, the proximal mapping $\prox_{\delta_\mathcal{E}}^\lambda(\xbf)$ takes a particularly simple form. From Equation \ref{eqn:indicator-function} and Definition \ref{def:prox-operator}, we can see that it equals the Euclidean projection operator $\mathcal{P}$ onto the set $\mathcal{E}$, i.e., 
\begin{eqnarray*}
\prox_{\delta_\mathcal{E}}^\lambda(\xbf) = \underset{\omegabf \in \mathcal{E}}{\arg\min}\; \frac{1}{2\lambda}\lVert \omegabf - \xbf \rVert^2 = \underset{\omegabf \in \mathcal{E}}{\arg\min}\; \lVert \omegabf - \xbf \rVert = \mathcal{P}_{\mathcal{E}}(\xbf) , \text{\,\, for all } \lambda > 0.
\end{eqnarray*}
Let $d_\mathcal{E}(\xbf)$ denote the Euclidean distance from the point $\xbf$ to the set $\mathcal{E}$, namely,
\begin{eqnarray*}
d_\mathcal{E}(\xbf) & = & \underset{\ybf \in \mathcal{E}}{\inf}\; \lVert \xbf - \ybf \rVert.
\end{eqnarray*}
Since $P_\mathcal{E}(\xbf)$ is the point in $\mathcal{E}$ that is closest in Euclidean distance to $\xbf$, 
\begin{eqnarray*}
d_\mathcal{E}(\xbf) & = & \lVert \xbf- \mathcal{P}_{\mathcal{E}}(\xbf)\rVert.
\end{eqnarray*}
Using Definition \ref{def:moreau-envelope}, the Moreau-Yosida envelope $\delta^\lambda_{\mathcal{E}}(\xbf)$ of $\delta_{\mathcal{E}}(\xbf)$ is
\begin{eqnarray*}
\delta^\lambda_{\mathcal{E}}(\xbf) = \frac{1}{2\lambda}\lVert \xbf- \mathcal{P}_{\mathcal{E}}(\xbf)\rVert^2 = \frac{1}{2\lambda}d^2_\mathcal{E}(\xbf).
\end{eqnarray*}

\subsection{Projections onto Epigraphs}

The key algorithmic primitive in our ProxMCMC framework is the projection onto the set $\mathcal{E}$. For regularized estimation problems, $\mathcal{E}$ is the epigraph of the regularization function $P(\thetabf)$, namely,
\begin{eqnarray*}
\mathcal{E} = \epi(P) = \{(\thetabf, \alpha) : P(\thetabf) \leq \alpha\}.
\end{eqnarray*}
Projection onto epigraphs is well known \citep{Beck17FOM} and is given by
\begin{eqnarray}\label{eqn:epiprojection}
\mathcal{P}_{\mathcal{E}}(\thetabf,\alpha) &=&\begin{cases}
(\thetabf,\alpha) & P(\thetabf)\le\alpha\\
\left(\prox^{\nu^*}_P(\thetabf),\alpha+\nu^* \right) & P(\thetabf)>\alpha
\end{cases},
\end{eqnarray}
where $\nu^*$ is any positive root of the auxiliary function $F(\nu)  = P\left(\operatorname{prox}^{\nu}_P(\thetabf)\right)-\nu-\alpha$, and can be found using bisection.

\section{An illustrative case study}
This section introduces our framework using a canonical example, the lasso regression \citep{Tibshirani96Lasso}. We have chosen the lasso because of its simplicity and familiarity to many readers, rather than as the motivation of this paper. The real power of ProxMCMC will be demonstrated on more complex models later. The lasso solves the following minimization problem, 
\begin{eqnarray}
\text{minimize} \quad \frac 12 \|\ybf - \Xbf \betabf\|_2^2 + \rho \|\betabf\|_1, \label{eqn:lasso-penalized-objective}
\end{eqnarray}
where $\ybf \in \Real^n$ is a vector of continuous responses, $\Xbf \in \Real^{n\times p}$ is a design matrix,  $\betabf\in \Real^p$ is the vector of regression coefficients, and $\rho$ is a nonnegative regularization strength parameter that trades off model fit with sparsity in the estimate of $\betabf$. 
To solve this problem in the ProxMCMC framework, we first write the regularized form \eqref{eqn:lasso-penalized-objective} in an equivalent constrained form
\begin{eqnarray*}
&\text{minimize}& \frac 12 \|\ybf - \Xbf \betabf\|_2^2 \\
&\text{subject to}& \|\betabf\|_1 \le \alpha,
\end{eqnarray*}
where the constraint parameter $\alpha$ is in one-to-one correspondence with the regularization strength parameter $\rho$. For this reason, we will also call $\alpha$ the regularization strength parameter. A Bayesian hierarchical model is specified for the constrained formulation of lasso:
\begin{itemize}
\item Data likelihood: $\Ybf \mid \betabf, \sigma^2 \sim N(\Xbf \betabf, \sigma^2 \Ibf)$,
\item A prior $\pi(\sigma^2)$ for the variance: $\sigma^2 \sim IG(r_{\sigma^2}, s_{\sigma^2})$, where $IG(r, s)$ denotes the Inverse-Gamma distribution with scale parameter $r$ and shape parameter $s$ (mean = $\frac{r}{s-1}$ for $s>1$),
\item A prior $\pi(\betabf \mid \alpha)$ for $\betabf$
conditional on $\alpha$, namely 
$$
\pi(\betabf \mid \alpha) = \frac{p! }{\alpha^p 2^p} \exp\left [-\delta_{\mathcal{E}}(\betabf, \alpha) \right],
$$
where $\mathcal{E} = \{(\betabf, \alpha) : \lVert \betabf \rVert_1 \leq \alpha \}$  and $\frac{p!}{\alpha^p 2^p}$ is the reciprocal of the volume of $\mathcal{E}$. Intuitively, $\pi(\betabf \mid \alpha)$ is a flat prior over an $\ell_1$-ball of radius $\alpha$.
\item A prior $\pi(\alpha)$ for the $\ell_1$-regularization strength parameter $\alpha$: $\alpha \sim IG(r_\alpha, s_\alpha)$.
\end{itemize}
The distribution $\pi(\betabf, \alpha) = \pi(\betabf \mid \alpha) \cdot \pi(\alpha)$ specifies a prior on the epigraph $\mathcal{E} = \{(\betabf, \alpha): \|\betabf\|_1 \le \alpha\} \subset \Real^{p+1}$. The posterior log-density, up to an irrelevant additive constant, is
\begin{equation*}
\begin{split}
&\log \pi(\betabf, \sigma^2, \alpha)\\
=& - \left(\frac{n}{2} + s_{\sigma^2} + 1\right) \log \sigma^2 - \frac{\|\ybf - \Xbf \betabf\|^2 + 2r_{\sigma^2}}{2\sigma^2}  \\
 & - (s_\alpha+p+1)\log \alpha - \frac{r_\alpha}{\alpha} - g(\betabf, \alpha),
\end{split}
\end{equation*}
where $g(\betabf,\alpha) = \delta_{\mathcal{E}}(\betabf, \alpha)$. Unfortunately, the posterior is not differentiable because it contains the non-differentiable indicator function $g(\betabf,\alpha)$. As a result, sampling algorithms for smooth log-densities cannot be directly applied. 

The key idea of the proposed ProxMCMC method is simple: find a smooth approximation to the non-differentiable posterior so it can be easily sampled from. Specifically, we approximate $g(\betabf,\alpha)$ with its Moreau-Yosida envelope $g^\lambda(\betabf, \alpha)$ and  substitute $g(\betabf,\alpha)$ with $g^\lambda(\betabf, \alpha)$ in the posterior. As mentioned in Section~\ref{sec:background}, $g^\lambda(\betabf, \alpha)$ approximates $g(\betabf,\alpha)$ arbitrarily well as the positive scaling constant $\lambda$ tends to $0$, so the smoothed posterior log-density
\begin{equation*}
\begin{split}
&\log \pi^\lambda(\betabf, \sigma^2, \alpha)\\
=& - \left(\frac{n}{2} + s_{\sigma^2} + 1\right) \log \sigma^2 - \frac{\|\ybf - \Xbf \betabf\|^2 + 2r_{\sigma^2}}{2\sigma^2}  \\
 & - (s_\alpha+p+1)\log \alpha - \frac{r_\alpha}{\alpha} - g^\lambda(\betabf, \alpha),
\end{split}
\end{equation*}
can be made arbitrarily close to $\log \pi(\betabf, \sigma^2, \alpha)$ as $\lambda$ tends to $0$. Since $\log \pi^\lambda(\betabf, \sigma^2, \alpha)$ is smooth, it can be readily sampled using any sampling algorithms for smooth log-densities. Hamiltonian Monte Carlo (HMC) \citep{Neal11HMC} is used in this paper due to its efficiency and generality. The last step of our algorithm is to log-transform non-negative parameters to make their domains unconstrained, which is a requirement for HMC. The smooth posterior under the parameterization $(\betabf, \log \sigma^2, \log \alpha)$ is
\begin{equation*}
\begin{split}
&\log \pi^\lambda(\betabf, \log \sigma^2, \log \alpha)\\
= &- \left(\frac{n}{2} + s_{\sigma^2}\right) \log \sigma^2-\frac{\|\ybf - \Xbf \betabf\|^2 + 2r_{\sigma^2}}{2\sigma^2} \\
& -(s_\alpha+p)\log \alpha - \frac{r_\alpha}{\alpha} - g^\lambda (\betabf,\alpha).
\end{split}
\end{equation*}

Before presenting numerical results, it is worth pausing to reflect on the power of the proposed method. Despite its simplicity, it demonstrates remarkable versatility as its extension beyond sparsity can be readily seen. To develop ProxMCMC algorithms for new regularized problems, one simply needs to find the corresponding Moreau-Yosida envelopes and proximal mappings, both of which are well-known for many non-smooth regularizations \citep{Beck17FOM}. The same idea can be applied to constrained problems in a similar manner, thus substantially broadening the range of problems that can be solved by ProxMCMC. Moreover, nothing prevents us from applying ProxMCMC to problems that encompass both regularizations and constraints. Additionally, the regularization strength parameter is seamlessly integrated into the inferential procedure in the proposed ProxMCMC method, rendering it fully Bayesian.

\begin{figure}[!tbp]
\centering
\includegraphics[width=0.8\textwidth]{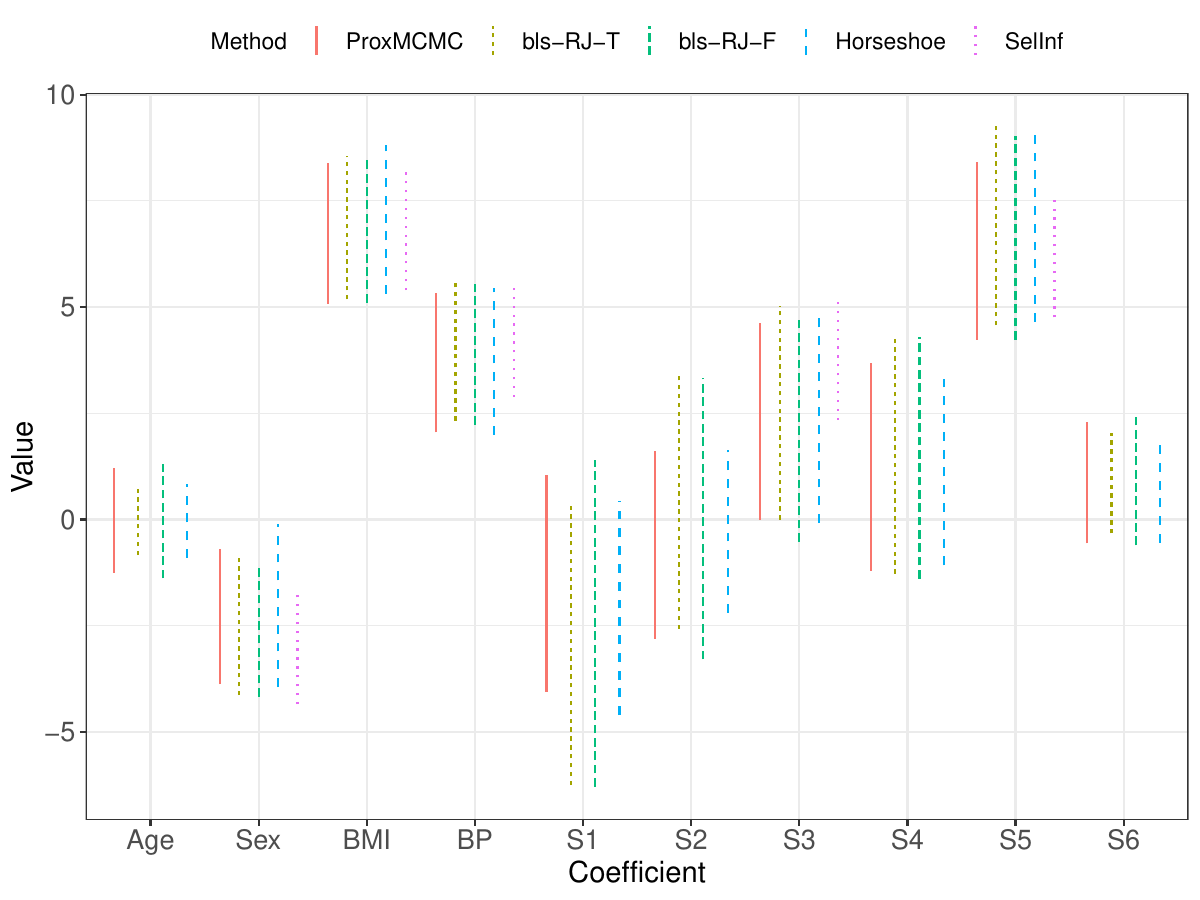}
  \caption{95\% credible intervals calculated by ProxMCMC , Bayesian lasso (bls), and horseshoe prior for the diabetes data set \citep{efron2004least}. Also shown are the 95\% selective inference (SelInf) confidence intervals for the five variables selected by lasso using 10-fold cross-validation. bls-RJ-T refers to Bayesian lasso with reversible jump MCMC (RJMCMC). bls-RJ-F indicates that RJMCMC is not used. } 
  \label{fig:blasso-proxmcmc}
\end{figure}

To see whether ProxMCMC gives reasonable results compared with existing methods such as Bayesian lasso and horseshoe prior, we apply them on the diabetes data set used by \citet{efron2004least}. The outcome is a quantitative measure of disease progression over a year, and the covariates are age, sex, body mass index, average blood pressure, and six blood serum measurements. All variables are standardized to have zero mean and unit variance. For Bayesian lasso, we use the \texttt{blasso} function from the \texttt{R} package \texttt{monomvn} \citep{gramacy2019monomvn} with default parameters. We show the results of Bayesian lasso with and without using reversible jump MCMC (RJMCMC) to perform model selection. For the horseshoe prior, we use the \texttt{R} package \texttt{horseshoe} \citep{stephanie2019horseshoe} and set function parameters \texttt{method.tau} and \texttt{method.sigma} to be \texttt{"truncatedCauchy"} and \texttt{"Jeffreys"}, respectively. For ProxMCMC, we set $\lambda = 0.001$, $\sigma^2 \sim IG(0.01, 0.01)$, and $\alpha \sim IG(1, 10+2)$. We also calculate the 95\% selective inference confidence intervals \citep{LeeSunSunTaylor2016PostSelectionInference} using the \texttt{R} package \texttt{selectiveInference} \citep{Tibshirani19Pkg}. Since selective inference requires a model to be selected first, we use lasso with 10-fold cross-validation and choose the largest regularization parameter such that the error is within 1 standard error of the minimum (the \texttt{lambda.1se} option from the \texttt{glmnet} package). Figure \ref{fig:blasso-proxmcmc} shows the 95\% interval estimates of the regression coefficients computed by each method.  We see that for null covariates, the credible intervals of Bayesian lasso are narrower when model selection by RJMCMC is used. This is because RJMCMC results in many exact zeros (75\% in this example) in the posterior sample, which reduces the width of credible intervals. When RJMCMC is not used, the credible intervals of the null covariates become wider and are similar to those obtained by ProxMCMC. The credible intervals from the horseshoe prior are narrower for null covariates, but for non-null covariates, the widths of the intervals are similar regardless of which method is used. 
The selective inference confidence intervals are calculated conditional on a selected model, and their coverage guarantee is in the frequentist sense, so they are not directly comparable with credible intervals. Nevertheless, we included them in the plot as a reference. 

\section{Methodology}\label{sec:methodology}

Having seen how to apply ProxMCMC to the special case of lasso, we next present the framework in greater generality. Our proposed ProxMCMC method consists of three steps.

\textbf{1. Likelihood and prior.} The first step is to specify a likelihood model for the data and priors for model parameters, which is a standard step in Bayesian modeling. Let $\taubf \in \Real^p$ denote parameters that are subject to regularizations or constraints, $\etabf \in \Real^q$ denote all other parameters including the regularization strength parameter $\alpha$, and $\thetabf = (\taubf^T, \etabf^T)^T \in \Real^d$ ($d = p + q$) denote all model parameters. Further let $\ell(\thetabf)$ be the log-likelihood and $\pi(\etabf)$ be the prior density for $\etabf$. The prior for $\taubf$ depends on whether the problem involves regularization, constraints, or both. 

For regularized problems, the prior for $\taubf$, conditional on the regularization strength parameter $\alpha$, is
$$\pi(\taubf\mid \alpha) = c \cdot \exp[-\delta_\mathcal{E}(\taubf, \alpha)],$$
where $c$ is a normalizing constant, and $\mathcal{E}$ is the epigraph of the regularization (penalty) function $P(\taubf)$, i.e., $\mathcal{E} = \epi(P) = \{(\taubf, \alpha) : P(\taubf) \leq \alpha \}$. For this reason we refer to $\pi(\taubf\mid \alpha)$ as the \emph{epigraph prior}. Since the regularization strength parameter $\alpha$ must be nonnegative, it requires a prior with nonnegative support. We find that placing an inverse Gamma prior on $\alpha$ works well in practice.

\begin{figure}[!tbp]
\centering
\includegraphics[width=0.49\textwidth]{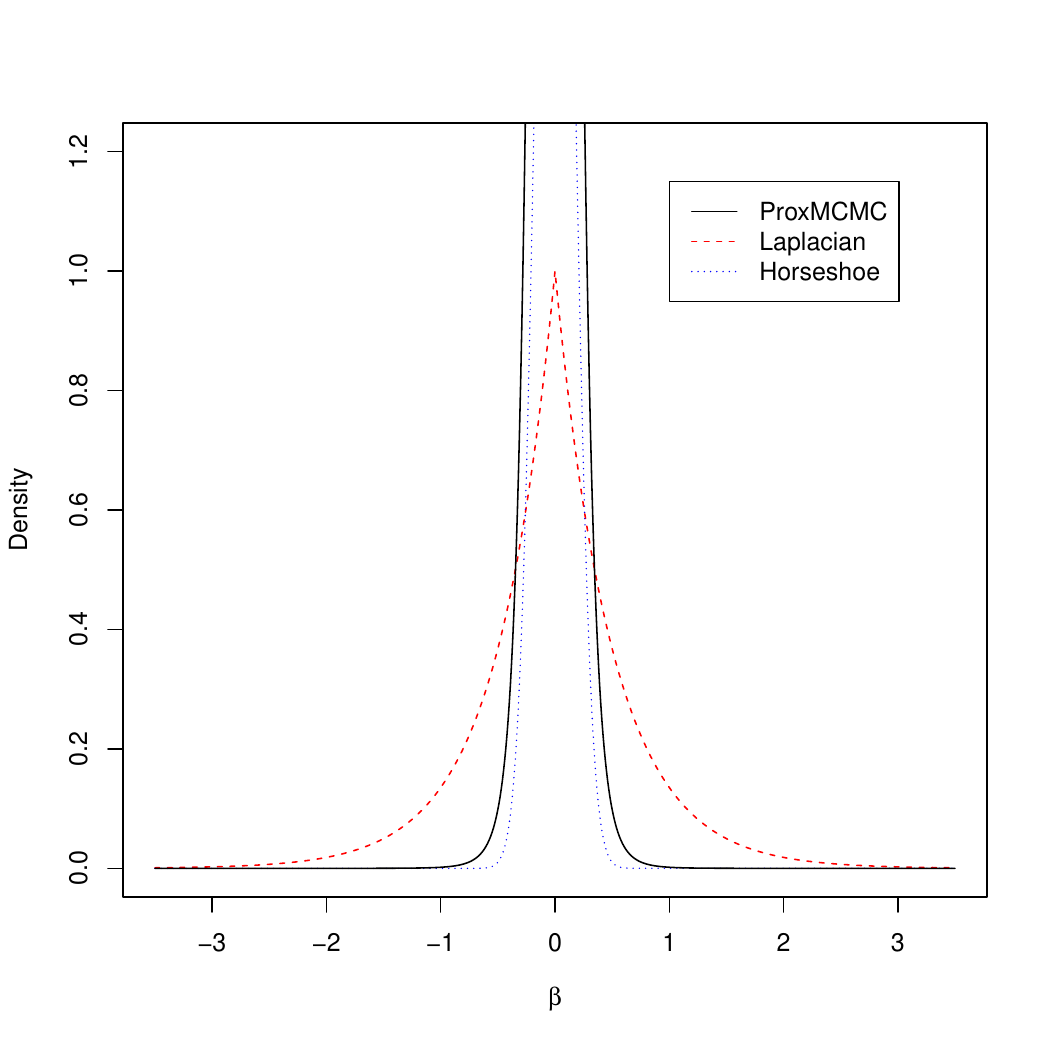}
  \caption{The density of ProxMCMC epigraph prior,  Laplacian prior, and horseshoe prior.}
  \label{fig:three-priors}
\end{figure}

To provide intuition on how the epigraph prior differs from existing alternatives, consider the simple case where a scalar parameter $\beta$ is regularized with the 
$\ell_1$-norm. The epigraph is
$\mathcal{E} = \{(\beta, \alpha): |\beta| \le \alpha\}$. With an $IG(r,s)$ prior on $\alpha$, the marginal density for $\beta$ is 
\begin{equation*}
f_\beta(t) = \int_{|t|}^\infty \frac{1}{2\alpha} \pi(\alpha)d\alpha = \frac{s}{2r}\big[1 - F_{IG(r, s+1)}(|t|)\big],
\end{equation*}
where $F_{IG(r, s+1)}(|t|)$ is the cumulative distribution function of $IG(r, s+1)$ evaluated at $|t|$. By comparing the ProxMCMC epigraph prior with  Laplacian prior and horseshoe prior, we can see from Figure~\ref{fig:three-priors} that it shrinks small $\beta$ while allowing strong signals to remain large. We would like to reiterate that the main motivation behind ProxMCMC is not to introduce yet another sparsity-inducing prior but rather to address problems that encompass constraints and more complex regularizations. The $\ell_1$-norm example is intended to offer intuition.

In the multivariate setting where more than one parameter is regularized, the ProxMCMC epigraph prior enforces negative correlation among components of $\betabf$. For example, it is clear from the lasso example, where the epigraph is given by $\mathcal{E} = \{(\betabf, \alpha): \|\betabf\|_1 \le \alpha\}$, that given $\alpha$, some components of $\betabf$ are forced to decrease as others take larger values. This repulsive feature distinguishes the ProxMCMC epigraph prior from other Bayesian priors such as the  Laplacian or horseshoe prior, where components of $\betabf$ are independent of each other conditional on the hyperparameter, and are marginally positively correlated.

For constrained problems, the set $\mathcal{E}$ refers to the constraint set instead of the epigraph, and the following prior for $\taubf$ is used: 
$$\pi(\taubf) = c \cdot \exp[-\delta_\mathcal{E}(\taubf)],$$
where $c$ is, again, a normalizing constant. 

For problems that encompass both regularization and constraints, two prior distributions are needed for $\taubf$: one to enforce the regularization and the other to enforce the constraints.

For simplicity of presentation, we will not distinguish between regularized problems and constrained problems, except in cases where distinction is necessary. We also abuse the notation slightly by using $g(\taubf)$ to denote either $\delta_\mathcal{E}(\taubf, \alpha)$ or $\delta_\mathcal{E}(\taubf)$, depending on the problem. 

Given the likelihood model and prior distributions, we have the posterior density
$$
\pi(\thetabf\mid Y) = \frac{e^{-U(\thetabf)}}{\int e^{-U(\sbf)}\, d\sbf} ,
$$
where $U(\thetabf) = f(\thetabf)+g(\taubf)$ and $f(\thetabf) = -\ell(\thetabf) - \log \pi(\etabf)$. The posterior $\pi(\thetabf\mid Y)$ is not differentiable because $g(\taubf)$ is not. By substituting $g(\taubf)$ with its Moreau-Yosida envelope $g^\lambda(\taubf)$, both $U^\lambda(\thetabf) = f(\thetabf) + g^\lambda(\taubf)$ and 
$$
\pi^\lambda(\thetabf\mid Y) = \frac{e^{-U^\lambda(\thetabf)}}{\int e^{-U^\lambda(\sbf)}\, d\sbf}
$$ 
become smooth functions. 

\textbf{2. Gradient.} The next step is to efficiently evaluate the gradient of the smoothed posterior log-density, which is another standard step in Bayesian modeling. For commonly used likelihood models and priors, the gradient can be computed numerically by auto-differentiation in software packages such as \texttt{Stan} \citep{STAN} and \texttt{Turing.jl} \citep{Ge18turing}.

As noted earlier, the existence of the gradient of the Moreau-Yosida envelope $g^{\lambda}(\taubf)$ depends on the convexity of the indicator function $g(\taubf)$, and thus on the convexity of the epigraph or the constraint set $\mathcal{E}$. When $g(\taubf)$ is convex, which is the case for many commonly used regularization and constraints, proximal mappings have been extensively studied in the optimization literature \citep{Beck17FOM}, and efficient implementations are available from mature libraries such as the FOM Matlab toolbox \citep{Beck19FOM}, the Python package \texttt{PyProximal}, and the Julia package \texttt{ProximalOperators.jl}.

When $g(\taubf)$ is non-convex, $g^\lambda(\taubf)$ is no longer differentiable. Under certain  regularity conditions, however, $g^\lambda(\taubf)$ is semidifferentiable and we can calculate a subgradient and use it in place of gradient in sampling algorithms. This approach will be demonstrated on the sparse low rank matrix regression example in Section \ref{sec:fixed-rank-matrix-regression}. 

\textbf{3. Sampling algorithm.} Finally, we invoke a gradient based sampling algorithm such as HMC or the Langevin algorithm to efficiently explore the posterior landscape. Software implementations include \texttt{DynamicHMC.jl}, \texttt{AdvancedHMC.jl}, and \texttt{pyhmc}, to name a few.

\vspace{0.5cm}

\noindent \textbf{Remark:} Before proceeding to examples, we pause to highlight ProxMCMC's close connection to distance majorization and proximal distance algorithms \citep{ChiZhouLange2014, XuChiLange2017, KeysZhouLange2019, LanderosLange2021, LanderosWuLange2022, Landeros2022}. Proximal distance algorithms are used to solve distance penalty problems of the form
\begin{eqnarray}
\label{eq:distance_majorization}
\text{minimize} \quad f(\thetabf) + \frac{\rho}{2} d_\mathcal{E}(\thetabf)^2,
\end{eqnarray}
where $f(\thetabf)$ is typically a negative log-likelihood term quantifying model fit, $\mathcal{E}$ is a target constraint set that we wish our estimate of $\thetabf$ to be close to, and $\rho$ is a nonnegative tuning parameter that trades off model fit with the amount of constraint violation quantified as the distance to $\mathcal{E}$. A solution to (\ref{eq:distance_majorization}) is a maximum a posteriori estimate under a distance-to-set prior $\pi(\thetabf) \propto \exp({-\frac{\rho}{2} d_\mathcal{E}(\thetabf)^2})$. Thus, the ProxMCMC method proposed here provides a fully Bayesian framework for generating posterior samples under a distance-to-epigraph set prior. Concurrent work in \citep{presman2022distance} uses distance-to-set priors to solve constrained Bayesian inference problems and discusses its advantages over prior literature on Bayesian constraint relaxation. 

\section{Examples}
The power of the proposed ProxMCMC method is illustrated on four examples, whose inference is either unknown or regarded as difficult. Since the potential applications of ProxMCMC are innumerable, our examples are not comprehensive. Nevertheless, we hope they serve as a starting point for readers to derive ProxMCMC algorithms for their own problems. See \citet{HengZhouChi23BayesianTrendFiltering} for an application of ProxMCMC to the Bayesian trend filtering problem.

\subsection{Constrained lasso} \label{sec:constrained-lasso}
Constrained lasso is a commonly used technique for analyzing compositional data and has been applied to problems such as consumer spending in economics, topic extraction of documents, and human microbiome analysis \citep{gaines2018algorithms,JamesPaulsonRusmevichientong20ConstrainedLasso}. The problem is formulated as 
\begin{eqnarray*}
\text{minimize} && \frac 12 \|\ybf-\Xbf\betabf\|_2^2 + \rho\|\betabf\|_1\\
\text{subject to} && \Abf\betabf = \bbf. 
\end{eqnarray*}
where $\ybf \in \Real^n$ is a vector of continuous responses, $\Xbf \in \Real^{n\times p}$ is a design matrix,  $\betabf\in \Real^p$ is the vector of regression coefficients, $\Abf$ and $\bbf$ impose constraints $\betabf$, and $\Abf$ has full row-rank. In compositional data analysis, for example, where each row of the design matrix $\Xbf$ represents proportions of a whole and sums to $1$, we can make $\betabf$ identifiable by constraining $\sum_i\betabf_i = 0$, which corresponds to $\Abf = \mathbf{1}_p^T$ (a row of $1$s) and $\bbf = 0$.

As in the lasso example, we use a normal likelihood model ($\Ybf \mid \betabf, \sigma^2 \sim N(\Xbf \betabf, \sigma^2 \Ibf)$) and inverse Gamma priors for $\sigma^2$ and $\alpha$ ($\sigma^2\sim IG(r_{\sigma^2},s_{\sigma^2}),\; \alpha\sim IG(r_\alpha,s_\alpha)$). Let $\mathcal{E}_1 = \{(\betabf, \alpha): \|\betabf\|_1 \le \alpha\}$ denote the epigraph of the $\ell_1$-norm and let $\mathcal{E}_2 = \{\betabf: \Abf\betabf = \bbf\}$ denote the constraint set. With the $(\betabf, \log \sigma^2, \log \alpha)$ parameterization, the smoothed posterior log-density  up to an irrelevant additive constant is
\begin{equation*}
\begin{split}
&\log \pi^\lambda(\betabf, \log \sigma^2, \log \alpha)\\
= &- \left(\frac{n}{2} + s_{\sigma^2}\right) \log \sigma^2-\frac{\|\ybf - \Xbf \betabf\|^2 + 2r_{\sigma^2}}{2\sigma^2} \\
& - s_\alpha \log \alpha - \frac{r_\alpha}{\alpha} - g_1^\lambda (\betabf, \alpha) - g_2^\lambda(\betabf),
\end{split}
\end{equation*}
where $g_1^\lambda(\betabf,\alpha)$ and $g_2^\lambda(\betabf)$ are the Moreau-Yosida envelopes of the indicator functions $g_1(\betabf,\alpha) = \delta_{\mathcal{E}_1}(\betabf, \alpha)$ and $g_2(\betabf) = \delta_{\mathcal{E}_2}(\betabf)$, respectively. From equation \eqref{eqn:epiprojection}, the proximal mapping of $g_1(\betabf,\alpha)$ is the projection onto the epigraph $\mathcal{E}_1$
\begin{eqnarray*}
\prox_{g_1}^{\lambda}(\betabf,\alpha) = \begin{cases}
(\betabf, \alpha) & \text{if $\|\betabf\|_1 \le \alpha$} \\
(S_{\nu^*}(\betabf), \alpha + \nu^*) & \text{if $\|\betabf\|_1 > \alpha$}
\end{cases},
\end{eqnarray*}
where $S_{\nu^*}$ is the soft-thresholding operator, the univariate form of which is given in \eqref{eqn:soft-thresholding}, and $\nu^*$ is any positive root of the nonincreasing function $\phi(\nu) = \|S_{\nu}(\betabf)\|_1 - \nu - \alpha$ \citep{Beck17FOM}.
The proximal mapping of $g_2$ is the projection onto the hyperplane given by 
\begin{eqnarray*}
\prox_{g_2}(\betabf) = \betabf - \Abf^T(\Abf\Abf^T)^{-1}(\Abf \betabf - \bbf).
\end{eqnarray*} 
The gradient of the posterior log-density is given block-wise by
\begin{eqnarray*}
\frac{\partial \log \pi^\lambda}{\partial \betabf} &=& \sigma^{-2} \Xbf^T(\ybf - \Xbf \betabf) - \lambda^{-1} \left[\betabf - \prox_{g_1}^\lambda(\betabf,\alpha)_{\betabf}\right] \\
& &-\lambda^{-1} \left[\betabf - \prox_{g_2}^\lambda(\betabf)\right] \\
\frac{\partial \log \pi^\lambda}{\partial \log \sigma^2} &=& - \left(\frac{n}{2} + s_{\sigma^2}\right) + \frac{\|\ybf - \Xbf \betabf\|^2 + 2r_{\sigma^2}}{2\sigma^2} \\
\frac{\partial \log \pi^\lambda}{\partial \log \alpha} &=& - s_\alpha + \frac{r_\alpha}{\alpha} - \lambda^{-1} \alpha [\alpha - \prox_g^\lambda(\betabf,\alpha)_{\alpha}].
\end{eqnarray*}
Numerical results will be presented in Section~\ref{sec:numerical-results}.

\subsection{Graphical lasso}

Given i.i.d. $p$-dimensional observations $\{\xbf_1, ..., \xbf_n\}$, where $\xbf_i \sim N(\mathbf{0}, \Sigmabf)$ and $\Sigmabf$ is a $p\times p$ covariance matrix, graphical lasso infers the underlying conditional dependency among covariates by estimating the precision matrix $\Thetabf = \Sigmabf^{-1}$ through maximizing the regularized log-likelihood
$$
-\frac n2 \tr(\Sbf\Thetabf) + \frac n2 \logdet(\Thetabf) - \rho\sum_{j \neq k}|\Thetabf_{jk}|,
$$
where $\Sbf$ is the sample covariance and $\rho$ is the regularization strength parameter.
Equivalently, we can maximize
$$
-\frac n2 \tr(\Sbf\Thetabf) + \frac n2 \logdet(\Thetabf) - g(\Thetabf,\alpha),
$$
where $g(\Thetabf,\alpha) = \delta_\mathcal{E}(\Thetabf,\alpha)$ and $\mathcal{E} = \{(\Thetabf, \alpha): \sum_{j \neq k}|\Thetabf_{jk}| \le \alpha\}$. The function $g(\Thetabf,\alpha)$ can be seen as the log-density (up to an additive constant) of the uniform prior for $\Thetabf$ over the $\ell_1$-ball $\{\Thetabf : \sum_{j \neq k}|\Thetabf_{jk}| \le \alpha\}$. With an  $IG(r_\alpha,s_\alpha)$ prior for $\alpha$, and after smoothing $g(\Thetabf,\alpha)$ with its Moreau-Yosida envelope $g^{\lambda}(\Thetabf,\alpha)$, the smoothed posterior log-density of $(\Thetabf, \log\alpha)$ is
\begin{equation*}
\begin{split}
\log \pi^\lambda(\Thetabf, \log\alpha) = &-\frac n2 \tr(\Sbf\Thetabf) + \frac{n}{2}\logdet(\Thetabf) \\
& - s_\alpha\log \alpha - \frac{r_\alpha}{\alpha} - g^{\lambda}(\Thetabf,\alpha).
\end{split}
\end{equation*}
Since HMC works on unconstrained domains, but $\Thetabf$ needs to be positive definite, we parameterize $\Thetabf$ in terms of its lower Cholesky factor $\Lbf$. Adjusting for the log-Jacobian terms, the smoothed posterior log-density becomes 
\begin{equation*}
\begin{split}
\log \pi^\lambda(\Lbf, \log\alpha) = &-\frac n2 \tr(\Sbf\Lbf\Lbf^T) + \frac{n}{2}\logdet(\Lbf\Lbf^T)\\
& -s_\alpha\log \alpha - \frac{r_\alpha}{\alpha} - g^{\lambda}(\Lbf\Lbf^T,\alpha) \\
&+ p\log(2) + \sum_{j=1}^p(p-j+2)\Lbf_{jj}.
\end{split}
\end{equation*}
The gradients are
\begin{eqnarray*}
\nabla_{\vech \Lbf} \log \pi^\lambda &=& -\big(n\vech (\Sbf\Lbf)\big)^T + n\big(\vech(\Lbf^{-1})^T \big)^T \\
&& -\frac 2\lambda \bigg(\vech\Big([ \Thetabf - \prox_g^\lambda(\Thetabf,\alpha)_{\Thetabf}]\Lbf\Big)\bigg)^T  \\
&&+\Big(\vech\big(\text{diag}(p+1, p, ..., 2)\big)\Big)^T\\
\frac{\partial \log \pi^\lambda}{\partial \log \alpha} &=& - s_\alpha + \frac{r_\alpha}{\alpha} - \lambda^{-1} \alpha [\alpha - \prox_g^\lambda(\Thetabf,\alpha)_{\alpha}],
\end{eqnarray*}
where $\vech(\Lbf)$ denotes the vector obtained from stacking the columns of the lower triangular part of the square matrix $\Lbf$.

\subsection{Matrix completion}
Given a matrix $\Ybf\in \Real^{n\times m}$ with entries only observed on the index set $\Omega=\{(i, j): y_{ij} \text{ is observed}\}$, \citet{Mazumder10SVDReg} proposed to complete the matrix by minimizing the convex objective function
$$
\frac 12 \|P_{\Omega}(\Ybf-\Xbf)\|_{\text{F}}^2 
+ \alpha \|\Xbf\|_*, 
$$
where $\Xbf$ is the completed matrix, $P_{\Omega}(\Ybf-\Xbf)$ is the projection of $\Ybf-\Xbf$ onto the set of observed entries $\Omega$, namely, the $(i,j)$-th entry of $P_{\Omega}(\Ybf-\Xbf)_{ij}$ is $y_{ij}-x_{ij}$ for $(i,j) \in \Omega$ and zero otherwise, $\alpha$ is the regularization strength parameter, and $\|\Xbf\|_*$ is the nuclear norm of $\Xbf$. The nuclear norm is defined as $\|\Xbf\|_*= \|\sigmabf(\Xbf)\|_1 = \sum_i \sigma_i(\Xbf)$, where $\sigma_1(\Xbf) \ge \cdots \ge \sigma_{m}(\Xbf) \ge 0$ are the singular values of $\Xbf$. To solve the matrix completion problem using ProxMCMC, we use the likelihood model $\vect (\Ybf) \sim N(\vect (\Xbf), \sigma^2 \Ibf)$, assume 
priors $\sigma^2 \sim IG(r_{\sigma^2},s_{\sigma^2})$ and $\alpha \sim IG(r_\alpha,s_\alpha)$, let $\mathcal{E} = \{(\Xbf, \alpha): \|\Xbf\|_* \le \alpha \}$ be the epigraph of $\|\cdot\|_*$, and let $g(\Xbf, \alpha) = \delta_\mathcal{E}(\Xbf, \alpha)$ be the corresponding indicator function. The smoothed posterior log-density using the $(\Xbf, \log \sigma^2, \log \alpha)$ parameterization is
\begin{equation*}
\begin{split}
&\log \pi^\lambda(\Xbf, \log \sigma^2, \log \alpha) \\
= &- \left(\frac{|\Omega|}{2} + s_{\sigma^2}\right) \log \sigma^2 -\frac{\sum_{(i,j) \in \Omega} (y_{ij}-x_{ij})^2 + 2r_{\sigma^2}}{2\sigma^2} \\
& -s_{\alpha} \log \alpha - \frac{r_\alpha}{\alpha} - g^\lambda (\Xbf,\alpha),
\end{split}
\end{equation*}
Let $\Xbf = \Ubf\Sigmabf\Xbf^T$ be the singular value decomposition of $\Xbf$, then the proximal mapping of $g(\Xbf, \alpha)$ is the projection given by
\begin{equation*}
\begin{split}
&\prox_g^{\lambda}(\Xbf,\alpha)\\ = &\begin{cases}
(\Xbf, \alpha) & \text{if $\|\Xbf\|_* \le \alpha$} \\
(\Ubf \text{diag}(S_{\nu^*}(\sigma(\Xbf)))\Vbf^T, \alpha + \nu^*) & \text{if $\|\Xbf\|_* > \alpha$}
\end{cases},
\end{split}
\end{equation*}
where $\nu^*$ is any positive root of the nonincreasing function $\phi(\nu) = \|S_{\nu}(\sigma(\Xbf))\|_1 - \nu - \alpha$. The gradient of the smoothed posterior log-density is 
\begin{eqnarray*}
\frac{\partial \log \pi^\lambda}{\partial \Xbf} &=& \sigma^{-2} \left[P_{\Omega}(\Ybf-\Xbf)\right]- \lambda^{-1} [\Xbf - \prox_g^\lambda(\Xbf,\alpha)_{\Xbf}], \\
\frac{\partial \log \pi^\lambda}{\partial \log \sigma^2} &=& - \left(\frac{|\Omega|}{2} + s_{\sigma^2}\right) + \frac{\sum_{(i,j) \in \Omega} (y_{ij}-x_{ij})^2 + 2r_{\sigma^2}}{2\sigma^2}, \\
\frac{\partial \log \pi^\lambda}{\partial \log \alpha} &=& - s_\alpha + \frac{r_\alpha}{\alpha} - \lambda^{-1} \alpha \left[\alpha - \prox_g^\lambda(\Xbf,\alpha)_{\alpha}\right].
\end{eqnarray*}

\subsection{Sparse low rank matrix regression} \label{sec:fixed-rank-matrix-regression}

We consider linear regression with matrix covariates, where the rank of the coefficient matrix is subject to regularization. One approach is to regularize the nuclear norm of the coefficient matrix \citep{zhou2014regularized}, for which the ProxMCMC algorithm is very similar to the matrix completion example above because they share the same proximal mapping. Alternatively, one can constrain the coefficient matrix to have a user-specified rank $k$ \citep{ZhouLiZhu13CPTensor}. Here we explore the second approach to illustrate the potential of ProxMCMC for problems where the regularization or constraints are not convex.

Let $y_i$ be the response of the $i$-th sample. Further let $\Zbf_i \in \Real^p$ and $\Xbf_i \in \Real^{q\times r}$  be the corresponding vector and matrix covariates, respectively. The model is 
$$
y_i = \Zbf_i^T\gammabf + \langle\Bbf, \Xbf_i\rangle + \epsilon_i,
$$ 
where $\gammabf$ and $\Bbf$ are the vector and matrix coefficients, $\langle\Bbf, \Xbf_i\rangle = \tr(\Bbf^T\Xbf_i) = \langle\vect\Bbf, \vect\Xbf_i\rangle$ is the inner product of the two matrices, and $\epsilon_i \sim N(0, \sigma^2)$. We fix $\rank(\Bbf)$ at a user-specified value $k$; the corresponding constraint set and indicator functions are $\mathcal{E}_1 = \{\Bbf: \rank(\Bbf) = k\}$ and $\delta_{\mathcal{E}_1}(\Bbf)$. To promote sparsity in $\Bbf$, we also incorporate an $\ell_1$-regularization on the entries of $\Bbf$; the epigraph set and indicator functions are $\mathcal{E}_2 = \{(\Bbf, \alpha): \|\vect\Bbf\|_1 \le \alpha\}$ and $\delta_{\mathcal{E}_2}(\Bbf,\alpha)$. With a flat prior for $\gammabf$ ($\pi(\gammabf) \propto 1$), an $IG(r_{\sigma^2},s_{\sigma^2})$ prior for $\sigma^2$, and an  $IG(r_\alpha,s_\alpha)$ prior for $\alpha$, the smoothed posterior log-density is
\begin{equation*}
\begin{split}
& \log \pi(\gamma, \Bbf, \log\sigma^2, \log\alpha) \\ = & - \frac{\sum_{i=1}^n(y_i - \Zbf_i^T\gamma - \langle\Bbf, \Xbf_i\rangle)^2 + 2r_{\sigma^2}}{2\sigma^2}\\
& -(\frac n2 + s_{\sigma^2})\log\sigma^2 - s_\alpha \log\alpha - \frac{r_\alpha}{\alpha} \\
&- g_1^\lambda(\Bbf) - g_2^\lambda(\Bbf, \alpha),
\end{split}
\end{equation*}
where $g_1^\lambda(\Bbf)$ and $g_2^\lambda(\Bbf,\alpha)$ are the Moreau-Yosida envelopes of $g_1(\Bbf) = \delta_{\mathcal{E}_1}(\Bbf)$ and $g_2(\Bbf,\alpha) = \delta_{\mathcal{E}_2}(\Bbf,\alpha)$, respectively. The proximal mapping of $g_1(\Bbf)$, given by the projection onto the set $\mathcal{E}_1$, can still be obtained relatively easily through thresholding the singular values of $\Bbf$. The gradient formula \eqref{eqn:moreau-gradient} for the Moreau-Yosida envelope, however, no longer holds because $g_1^\lambda(\Bbf)$ is not convex. The solution we explore below resorts to the subsmoothness property of Moreau-Yosida envelopes, for which we need the following definitions \citep{rockafellar2009variational}.

\begin{definition}
(\textbf{Prox-boundedness}) A function $g: \Real^n \rightarrow \bar{\Real}$ is prox-bounded if there exists $\lambda > 0$ such that its Moreau-Yosida envelope $g^\lambda > -\infty$ for some $\xbf\in \mathbb{R}^n$. The supremum of the set of all such $\lambda$ is the threshold $\lambda_g$ of prox-boundedness for $g$. 
\end{definition}
In the ProxMCMC framework, we only need the Moreau-Yosida envelope of indicator functions, for which we have $g^\lambda(\xbf) > -\infty$ for any $\lambda > 0$, so they are always prox-bounded and the threshold $\lambda_g = \infty$. 

\begin{definition}
(\textbf{Semidifferentiability}) Let $g: \mathbb{R}^n \rightarrow \bar{\mathbb{R}}$ and $\bar{\xbf}$ be a point such that $g(\bar{\xbf})$ is finite. If the (possibly infinite) limit $$\lim_{\tau \downarrow 0, \wbf' \rightarrow \wbf} \frac{g(\bar{\xbf} + \tau \wbf') - g(\bar{\xbf})}{\tau}$$ exists, it is the semiderivative of $g$ at $\bar{\xbf}$ for $\wbf$, and $g$ is  semidifferentiable at $\bar{\xbf}$ for $\wbf$. If this holds for every $\wbf$, $g$ is semidifferentiable at $\bar{\xbf}$.
\end{definition}

By \citet[Example 10.32]{rockafellar2009variational}, if $g(\xbf)$ is lower-semicontinuous, proper, and prox-bounded with threshold $\lambda_g$, then for $\lambda \in (0, \lambda_g)$, the Moreau-Yosida envelope $g^\lambda(\xbf)$ is 
semidifferentiable and the subgradient set is
$$
\partial g^\lambda(\xbf) \subset \lambda^{-1}\left[\xbf-\prox_g^\lambda(\xbf)\right].
$$

The function $g_1(\Bbf) = \delta_{\mathcal{E}_1}(\Bbf)$ satisfies the above conditions, so we can calculate its subgradient using the above formula and use it in place of the gradient in HMC.
\begin{eqnarray*}
\frac{\partial \log \pi}{\partial \gammabf} &=&  \sigma^{-2} \sum_i (y_i - \Zbf_i^T\gammabf - \langle\Bbf, \Xbf_i\rangle)Z_i, \\
\frac{\partial \log \pi}{\partial \Bbf} &=&  \sigma^{-2} \sum_i (y_i - \Zbf_i^T\gammabf - \langle\Bbf, \Xbf_i\rangle)X_i \\
&& -\lambda^{-1} \left[\Bbf - \prox_{g_1}^\lambda(\Bbf)\right] \\
&& - \lambda^{-1} \left[\Bbf - \prox_{g_2}^\lambda(\Bbf,\alpha)_{\Bbf}\right], \\
\frac{\partial \log \pi}{\partial \log \sigma^2}&=& - \left(\frac n2 + s_{\sigma^2}\right)\\
&& + \frac{\sum_{i=1}^n(y_i - \Zbf_i^T\gammabf - \langle\Bbf, \Xbf_i\rangle)^2 + 2r_{\sigma^2}}{2\sigma^2}, \\
\frac{\partial \log \pi}{\partial \log \alpha} &=& - s_\alpha + \frac{r_\alpha}{\alpha} - \lambda^{-1} \alpha \left[\alpha - \prox_{g_2}^\lambda(\Bbf,\alpha)_{\alpha}\right].
\end{eqnarray*}
Since $g^\lambda_1(\Bbf)$ is non-convex, $\prox_{g_1}^\lambda(\Bbf)$ is not unique. Our approach is to pick an arbitrary element in the proximal map set, which works well in practice.

\section{Numerical Results}\label{sec:numerical-results}
This section demonstrates the proposed ProxMCMC method through either simulation experiments or analysis of publicly available data sets.  

\subsection{Constrained lasso: simulated microbiome data}
\begin{figure}
    \includegraphics[width=0.3\linewidth]{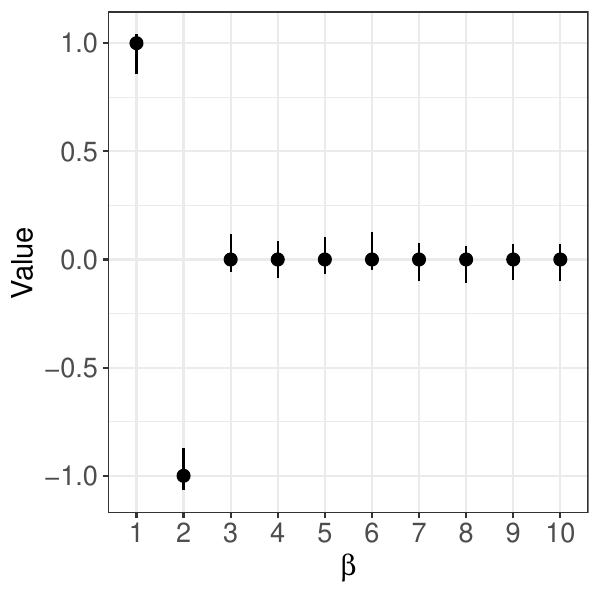}
    \includegraphics[width=0.3\linewidth]{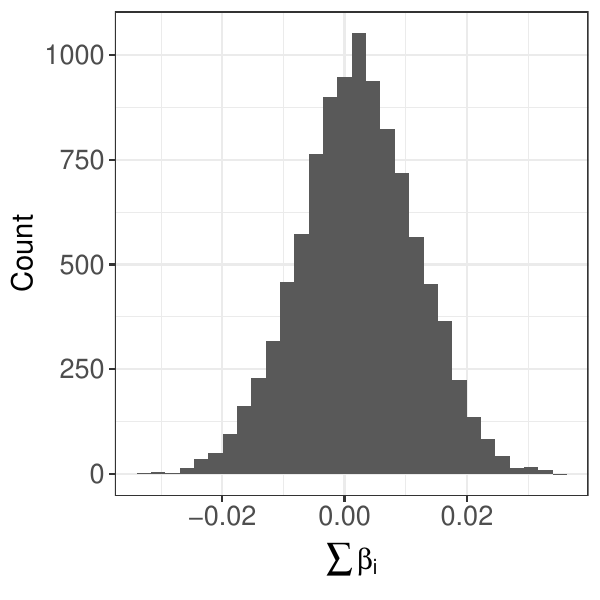}
    \includegraphics[width=0.3\linewidth]{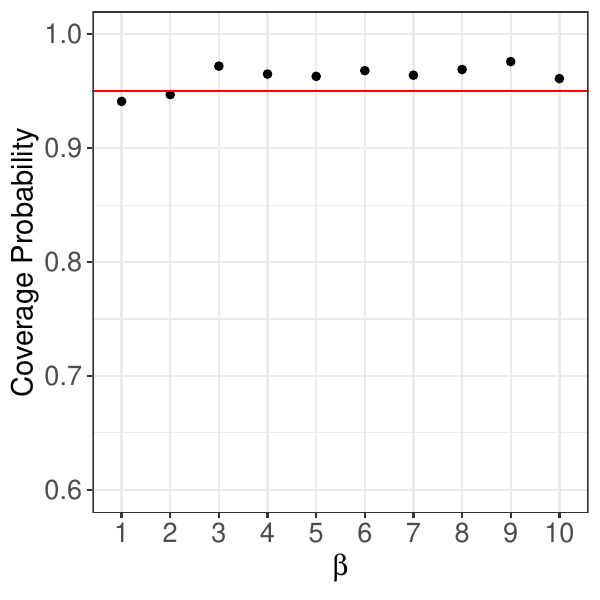}
    \caption{Left: 95\% credible intervals for constrained lasso models parameters from one simulated data set. Dots mark the truth. Middle: Histogram of $\sum_i \beta_i$ over $10,000$ samples for the same data set used on the left. Right: Coverage probability for model coefficients calculated from $1,000$ simulated data sets. The red line is the nominal level of 95\%.} \label{fig:constrained-lasso}
\end{figure}

We illustrate the ProxMCMC method for constrained lasso using a simulated microbiome data set. The 16S microbiome sequencing technology measures the number of various organisms called operational taxonomic units (OTUs) in a biological sample. For statistical analysis, counts are normalized into proportions for each sample, resulting in a design matrix $\Xbf$ where each row sums to 1, which makes it necessary to constrain regression parameters so that they are identifiable. We use the popular sum-to-zero constraint ($\sum_j \beta_j = 0$) in this example. We set sample size $n = 1000$ and number of OTUs $p=10$. The design matrix $\Xbf$ is generated as follows. First, each entry in $\Xbf$ is sampled i.i.d.\@ from a uniform distribution ($U_{[0,1]}$). Second, the rows of $\Xbf$ are scaled so that each row sums to 1. We set $\beta_1 = 1$, $\beta_2=-1$ and the remaining $\beta_j$ to 0 so that 20\% of the entries in $\betabf$ are nonzero. The noise is generated from a normal distribution with mean 0 and $\sigma = 0.1$ so that the sample signal-to-noise ratio $\Var(\Xbf\betabf)/\sigma^2$ is approximately $0.7$. We use $IG(0.01, 0.01)$ as a prior for $\sigma^2$ and $IG(1, p+1)$ as a prior for $\alpha$, set $\lambda = 10^{-5}$, and ran HMC for 10,000 iterations. The experiment is repeated 1000 times to estimate the coverage probability. Figure \ref{fig:constrained-lasso} (left) shows the 95\% credible intervals and the true values (black dots) for the regression parameters for the first simulated data set. We can see that credible intervals provide good coverage of the truth. Figure \ref{fig:constrained-lasso} (middle) shows the histogram of $\sum_j \beta_j$ for posterior samples from the first simulated data set. The histogram is highly concentrated around 0, which shows that the posterior samples satisfy the sum-to-zero constraint well. To measure the sampling efficiency of our algorithms, we calculate the effective sample size of the slowest moving component of the multivariate posterior samples. The slowest moving component can be obtained by first performing a principal components analysis on the posterior covariance matrix and then projecting the posterior samples onto the most prominent eigenvector \citep{durmus2018efficient}. After obtaining the slowest moving component, which is a vector of the same length as the number of posterior samples, we can calculate its effective sample size with the \texttt{ess\_rhat} function from the \texttt{MCMCDiagnosticTools.jl} package. Using this method, the effective sample size of the slowest $\betabf$ component is 7044.  Finally, Figure~\ref{fig:constrained-lasso} (right) shows the coverage probability of model parameters. Results indicate that the coverage probability of ProxMCMC credible intervals are very close to the nominal level of 95\%.  

\subsection{Graphical lasso: cytometry data}

\begin{figure}[!tbp]
\centering
\includegraphics[width=0.46\textwidth]{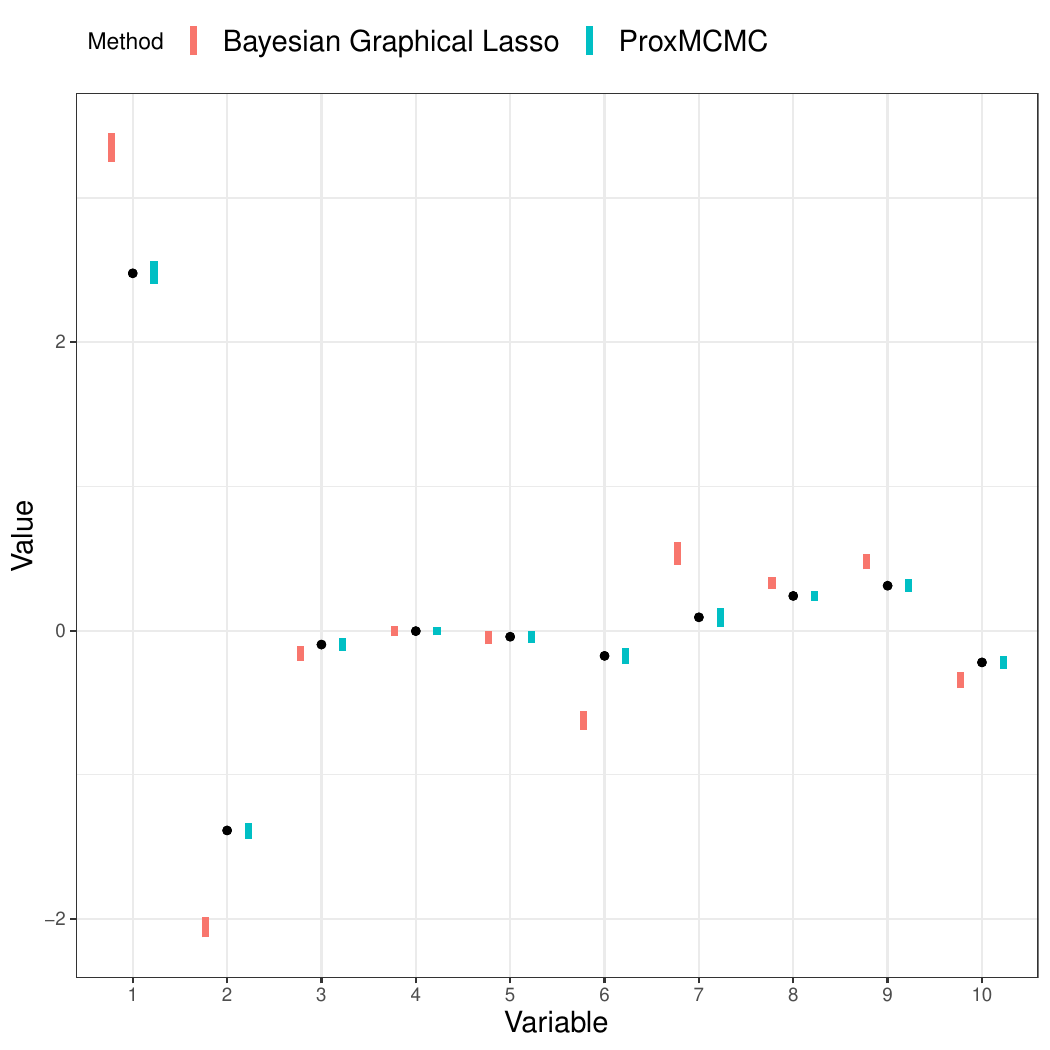}
  \caption{Comparing the 95\% credible intervals of Bayesian graphical lasso versus ProxMCMC on the cytometry data. Black dots are estimates obtained from 5-fold cross-validated graphical lasso.}
  \label{fig:graphical-lasso}
\end{figure}

We compare ProxMCMC with Bayesian graphical lasso \citep{wang2012bayesian} on the cell-signalling data from \citet{sachs2005causal}, which was used in the original graphical lasso paper \citep{friedman2008sparse}. The data set contains flow cytometry measurements on $p = 11$ proteins and $n = 7466$ cells. We first use the \texttt{R} package \texttt{CVglasso} to compute 5-fold cross-validated graphical lasso estimates for $\Thetabf$, which are used as references for the comparison between ProxMCMC and Bayesian graphical lasso. For Bayesian graphical lasso, we use the \texttt{R} package \texttt{BayesianGLasso} \citep{wang2012bayesian}. We experimented with both the default prior and other prior settings but found little difference, so we report the results using the default prior (Gamma distribution with shape parameter $1$ and scale parameter $0.1$). For ProxMCMC, we use an $IG(1, p+1)$ prior for $\alpha$ and set $\lambda = 0.01$.  We ran 10,000 iterations for both methods. Figure \ref{fig:graphical-lasso} displays the 95\% credible intervals. Due to the large number of parameters, we only show the results for the first ten parameters in the plot, but the same pattern is observed for other parameters. We can see that ProxMCMC credible intervals are consistently narrower and provide good coverage of the graphical lasso estimates, whereas those provided by Bayesian graphical lasso can be wide or fail to cover the cross-validated estimates.  Among all 66 parameters, all ProxMCMC credible intervals cover the reference values whereas only 24\% of Bayesian graphical lasso credible intervals do.  The effective sample size of the slowest $\Thetabf$ component is $5,540$.

\subsection{Matrix completion: simulated matrix}
We simulate the true low-rank matrix as $\Ybf = \Xbf_1\Xbf_2 + \sigma\Ebf$, where $\Xbf_1\in\Real^{50\times 2}$ , $\Xbf_2\in \Real^{2\times 50}$, $\sigma = 0.5$, and entries of $\Xbf_1,\Xbf_2,\Ebf$ are generated from the standard normal distribution. We randomly mask 25\%, 50\%, and 75\% of the entries and apply ProxMCMC to calculate the posterior median and 95\% credible intervals for the missing entries. We use an $IG(0.01, 0.01)$ prior for $\sigma^2$ and an $IG(1, 50\times 50+1)$ prior for $\alpha$, and set $\lambda = 0.01$. The number of HMC samples is set at 1000. For comparison, we also try an empirical Bayesian method called the stochastic approximation proximal gradient (SAPG) \citep{de2020maximum,vidal2020maximum}, and use the SK-ROCK method \citep{pereyra2020accelerating} for posterior sampling. The details of this approach is left to the supplementary materials. Table \ref{table:mc} displays the mean absolute deviation (MAD) averaged over missing entries and the percentage of missing entries covered by their 95\% credible intervals for the two methods at different missing rate. As expected, the posterior average MAD increases as the missing rate increases. We also see that, for a given missing rate, the average MAD of ProxMCMC is lower than that of SAPG, and the credible intervals provided by ProxMCMC cover an equal or higher percentage of missing entries than that provided by SAPG. The results indicate that ProxMCMC has superior statistical precision, likely because ProxMCMC is fully Bayesian and accounts for the uncertainty of $\alpha$ and $\sigma^2$, while SAPG commits to a single point estimate of $\alpha$ and $\sigma^2$ after hyperparameter calibration. We also emphasize that it is not straightforward to apply SAPG to problems with constraints, such as the constrained lasso or the sparse low rank matrix regression problem. Therefore ProxMCMC offers greater flexibility in model formulation. 

\begin{table}[ht]
\begin{center}
\begin{adjustbox}{width=0.8\textwidth}
\begin{tabular}{ccccc}
\toprule
\multicolumn{1}{c}{\,} & \multicolumn{2}{c}{ProxMCMC} & \multicolumn{2}{c}{SAPG}\\ \cmidrule(lr){2-3}\cmidrule(lr){4-5}
Percent Missing & Average MAD & Percent Covered & Average MAD & Percent Covered \\ \midrule
25\% & 0.23 & 100\% & 0.24 & 100\%\\
50\% & 0.34 & 100\%  & 0.42 & 99\%\\
75\% & 0.74 & 97\% & 0.79 & 93\%\\
\bottomrule
\end{tabular}
\end{adjustbox}
\end{center}
\caption{Comparison between ProxMCMC and stochastic approximation proximal gradient (SAPG) for the matrix completion example. ``Average MAD" is the mean absolute deviation averaged over missing entries; ``Percent covered" is the percentage of missing entries covered by their 95\% credible intervals.}
\label{table:mc}
\end{table}

\subsection{Sparse low rank matrix regression: detecting the butterfly signal}
\begin{figure}
\centering
\includegraphics[width=0.3\linewidth]{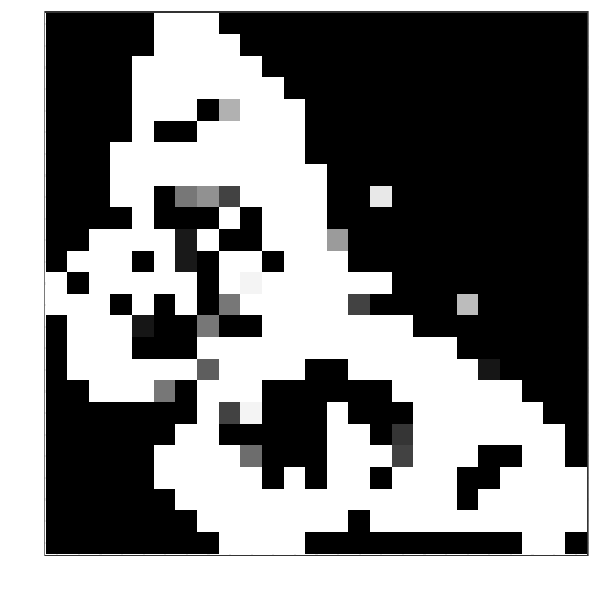}
\includegraphics[width=0.3\linewidth]{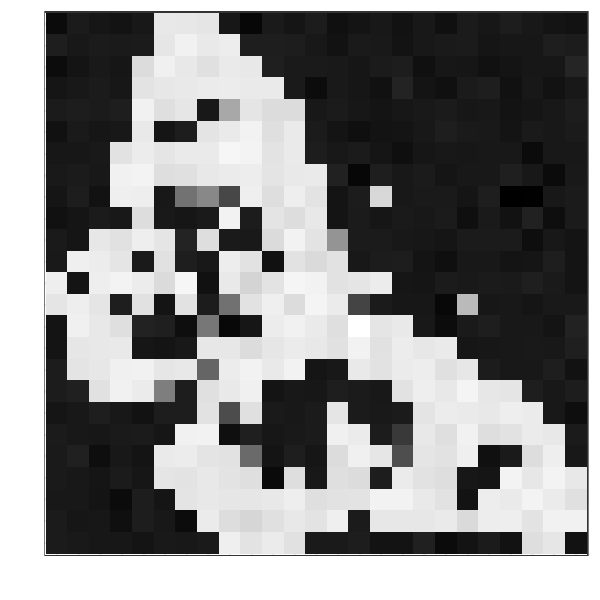}
\includegraphics[width=0.3\linewidth]{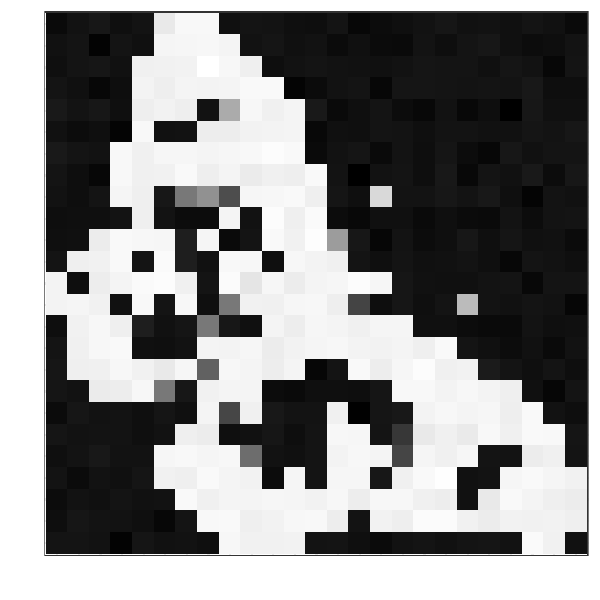}
\caption{ProxMCMC for sparse low rank matrix regression on the butterfly signal. Left: true signal; Middle: posterior mean at sample size 2000; Right: posterior mean at sample size 5000. Black pixels equal 0, white pixels 1, and grey pixels between 0 and 1.} \label{fig:SLMR}
\end{figure}

We simulate data from the following model: the mean response for the $i$-th sample is $\mu_i = \Zbf_i^T\gammabf + \langle \Bbf, \Xbf_i \rangle$, where $\Zbf_i\in \mathbb{R}^2$ and $\Xbf_i \in \mathbb{R}^{25 \times 25}$ are vector and matrix covariates, whose entries are generated from i.i.d. standard normal. We set the true $\gammabf = (1, 1)^T$ and let $\Bbf$ be the $25\times 25$ butterfly signal shown in Figure~\ref{fig:SLMR} (left), where black pixels equal 0, white pixels 1, and grey pixels between 0 and 1. The response for the $i$-th sample, $y_i$, equals $\mu_i + \epsilon_i$, where $\epsilon_i$ is generated from i.i.d. standard normal. We use an $IG(0.01, 0.01)$ prior for $\sigma^2$ and an $IG(\sum_i \sigma(\Bbf_0)_i, 2)$ prior for $\alpha$, where $\sigma(\Bbf_0)_i$ is the $i$-th singular value of $\Bbf_0$, and $\Bbf_0$ is the least squares estimate of $\Bbf$ obtained without regularization or constraints. We set the Moreau-Yosida envelope parameter $\lambda = 0.001$. Figure \ref{fig:SLMR} shows the true signal $\Bbf$ (left) and the posterior mean from 10,000 HMC samples at sample size $N=2000$ (middle) and $N=5000$ (right). For inference, we calculated the 95\% credible intervals for entries of $\Bbf$ and found that among the 625 ($=25\times 25$) entries, 94\% are covered by their 95\%  credible intervals at both sample sizes. The effective sample size of the slowest component of $\Bbf$ is 361 at $N=2000$, and 2054 at $N=5000$. 

\section{Discussion}

The examples above demonstrate that the ProxMCMC method is a highly flexible tool for obtaining statistical inference on regularized or constrained statistical learning problems. We find that it works well when the regularization or constraints are non-smooth and even non-convex.  In addition, by adopting epigraph priors, our method is fully Bayesian, eliminating the need for tuning the regularization strength parameter.

The Moreau-Yosida envelope parameter $\lambda$ controls how well the smoothed posterior approximates the original posterior. For constrained problems, a smaller $\lambda$ leads to better satisfaction of the constraints. For example, the histogram of $\sum_j \betabf_j$ from the constrained lasso simulation experiment is more concentrated around 0 when $\lambda$ is smaller. Choosing $\lambda$ values that are too small, however, renders slow mixing of the sampling algorithm. We leave a more in-depth investigation of this phenomenon to future work. For practical purposes, we recommend using smaller $\lambda$ when computational resources allow. Setting $\lambda = 0.001$ seems to work well in most applications as the examples show.

Finally, we emphasize that the four examples are meant to whet readers' appetites, not to satiate them. As demonstrated through these examples, the proposed ProxMCMC method is highly modular and can be readily extended to other problems. We hope that this paper offers sufficient detail for readers to explore new applications of the ProxMCMC algorithm.

\clearpage

\section*{Supplementary Material}
\setcounter{page}{1}
\setcounter{section}{0}
\setcounter{table}{0}
\setcounter{figure}{0}

\renewcommand{\thesection}{S.\arabic{section}}
\renewcommand{\thetable}{S.\arabic{table}}
\renewcommand{\theequation}{S.\arabic{equation}}
\renewcommand{\thefigure}{S.\arabic{figure}}

\section{Theoretical properties}

This section presents theoretical results for the ProxMCMC method. Our proofs, compared to that of \citet{durmus2018efficient}, extend to non-convex settings while \citep{durmus2018efficient} assumes convexity, and are simpler because we focus on the Moreau-Yosida envelope of indicator functions. As defined in Section 3 of the main text, $\thetabf \in \mathbb{R}^d$ represents all model parameters that include both the constrained or regularized parameters $\taubf \in \mathbb{R}^p$ and other parameters $\etabf \in \mathbb{R}^q$. We also use $\ell(\thetabf)$ for the log-likelihood and $\pi(\etabf)$ for the prior density of $\etabf$. In this section, for simplicity we assume that there is only one Moreau envelope $g^\lambda(\taubf)$. The argument for multiple Moreau envelopes will be nearly identical. Our main theoretical results are summarized as follows:

\begin{proposition}
\begin{enumerate}[label = (\arabic*)]
\item For any $\lambda > 0$, the smoothed posterior $\pi^\lambda(\thetabf \mid Y)$ defines a proper density of a probability measure on $\mathbb{R}^d$, i.e. $$0 < \int_{\mathbb{R}^d}e^{-U^\lambda(\thetabf)}\, d\thetabf < \infty.$$
\item Denote the maximum-a-posteriori (MAP) estimates of $\pi(\thetabf\mid Y)$ and $\pi^\lambda(\thetabf\mid Y)$ as $\hat{\thetabf}$ and $\hat{\thetabf}_\lambda$. For any sequence of $\{\lambda_k\}$ that monotonously decreases to 0, all limit points of the sequence $\{\hat{\thetabf}_{\lambda_k}\}$ are MAP estimates of $\pi(\thetabf\mid Y)$.
\item If $\pi(\thetabf\mid Y)$ defines a proper density on $\mathbb{R}^d$, i.e., $0 < \int_{\mathbb{R}^d}e^{-U(\thetabf)}\, d\thetabf < \infty$, then the approximation $\pi^\lambda(\thetabf \mid Y)$ converges to $\pi(\thetabf \mid Y)$ in total-variation as $\lambda \downarrow 0$, i.e., $$\lim_{\lambda \downarrow 0} \|\pi^\lambda(\thetabf \mid Y) - \pi(\thetabf \mid Y)\|_{\text{TV}} = 0.$$
\end{enumerate}
\end{proposition}

\begin{proof}
(Posterior properness) The properness of the smoothed posterior $\pi^\lambda(\thetabf \mid Y)$ follows from the fact that the Moreau-Yosida envelope of an indicator function is always nonnegative. Specifically, when $g = \delta_\mathcal{E}(\taubf)$, 
$$
g^\lambda(\taubf) = \frac{1}{2\lambda}d_\mathcal{E}(\taubf)^2 \geq 0,
$$ 
where $d_\mathcal{E}(\taubf) = \inf_{\ybf\in \mathcal{E}}d(\taubf,\ybf)$ is the distance from $\taubf$ to $\mathcal{E}$, so $-U^\lambda(\thetabf) = -f(\thetabf)- g^\lambda(\taubf) \leq -f(\thetabf)$, from which we have
$$
e^{-U^\lambda(\thetabf)} \leq e^{-f(\thetabf)} . 
$$
Since $f(\thetabf) = -\ell(\thetabf) - \log \pi(\etabf)$ and both the likelihood and the priors $\pi(\etabf)$ are integrable (note that $\etabf$ does not include constrained parameters), we have the desired result.

(Convergence of MAP) From the definition of $\hat{\thetabf}_{\lambda_k}$, we have
$$
\log\pi_{\lambda_k}(\hat{\thetabf}\mid Y) \le \log\pi_{\lambda_k}(\hat{\thetabf}_{\lambda_k}\mid Y).
$$
This is equivalent to
$$
-f(\hat{\thetabf}) \le -f(\hat{\thetabf}_{\lambda_k}) - g^{\lambda_k}(\hat{\taubf}_{\lambda_k}),
$$
because $\hat{\thetabf}$ satisfies the constraints. 
Then we have
$$
0 \le d_\mathcal{E}(\hat{\taubf}_{\lambda_k})^2\le 2\lambda_k (f(\hat{\thetabf})-f(\hat{\thetabf}_{\lambda_k})),
$$
which implies $d_\mathcal{E}(\hat{\taubf}_{\lambda_k})^2\downarrow 0$. We also have
$$
-f(\hat{\thetabf})\le \underset{k\rightarrow \infty}{\operatorname{liminf}}\;( -f(\hat{\thetabf}_{\lambda_k}) - g^{\lambda_k}(\hat{\taubf}_{\lambda_k}))\le \underset{k\rightarrow \infty}{\operatorname{limsup}}\;( -f(\hat{\thetabf}_{\lambda_k}) - g^{\lambda_k}(\hat{\taubf}_{\lambda_k})) \le \underset{k\rightarrow \infty}{\operatorname{limsup}}\; (-f(\hat{\thetabf}_{\lambda_k})).
$$
Thus for any limit point $\thetabf^*$ of $\{\hat{\thetabf}_{\lambda_k}\}$, we have $\taubf^*\in \mathcal{E}$ since $d_\mathcal{E}(\hat{\taubf}_{\lambda_k})^2\downarrow 0$. Also, $-f(\thetabf^*)\ge -f(\hat{\thetabf})$ due to the above inequality. Therefore, $\thetabf^*$ is a MAP estimate of $\pi(\thetabf\mid Y)$.

(Convergence in total-variation) 
Let $c = \int e^{-U(\sbf)}\, d\sbf$ and $c_\lambda = \int e^{-U^\lambda(\sbf)}\, d\sbf$. Since $g^\lambda(\xbf)$ uniformly bounds $g(\xbf)$ from below, i.e., $g^\lambda(\xbf) \leq g(\xbf)$ for all $\xbf$ \citep{rockafellar2009variational}, we have $U^\lambda(\xbf) \leq U(\xbf)$ and thus $c_\lambda \geq c$. Note that 
\begin{equation*}
\begin{split}
&\|\pi^\lambda - \pi\|_{\text{TV}} \\
=& \int |\pi^\lambda(\xbf) - \pi(\xbf)|\, d\xbf\\
=&\int_{\pi^\lambda \geq \pi} \left[\pi^\lambda(\xbf) - \pi(\xbf)\right]\, d\xbf\\
&+\int_{\pi^\lambda < \pi} \left[\pi(\xbf)- \pi^\lambda(\xbf)\right]\, d\xbf.
\end{split}
\end{equation*}
Let $\mathcal{A}_1 = \left\{\xbf: \pi^\lambda \geq \pi\right\}$ and $\mathcal{A}_2 = \left\{\xbf: \pi^\lambda < \pi\right\}$,
\begin{equation*}
\begin{split}
&\int_{\mathcal{A}_1} \left [\pi^\lambda(\xbf) - \pi(\xbf)\right]\, d\xbf\\
= &\int_{\mathcal{A}_1} \pi^\lambda(\xbf)\left[1 - \frac{\pi(\xbf)}{\pi^\lambda(\xbf)}\right]\, d\xbf\\
= &\int_{\mathcal{A}_1} \pi^\lambda(\xbf)\left[1 -\frac{c_\lambda}{c}e^{g^\lambda(\xbf) - g(\xbf)}\right]\, d\xbf\\
\leq & \int_{\mathcal{A}_1} \left[\pi^\lambda(\xbf) - e^{g^\lambda(\xbf) - g(\xbf)}\pi^\lambda(\xbf)\right]\, d\xbf\\
\leq &\,\, 1-\frac{c}{c_\lambda},
\end{split}
\end{equation*}
and 
\begin{equation*}
\begin{split}
&\int_{\mathcal{A}_2} \left [\pi(\xbf)-\pi^\lambda(\xbf)\right]\, d\xbf\\
= &\int_{\mathcal{A}_2} \pi(\xbf) \left[1 - \frac{\pi^\lambda(\xbf)}{\pi(\xbf)}\right]\, d\xbf\\
= &\int_{\mathcal{A}_2} \pi(\xbf)\left[1 -\frac{c}{c_\lambda}e^{g(\xbf) - g^\lambda(\xbf)}\right]\, d\xbf\\
\leq & \int_{\mathcal{A}_2} \pi(\xbf)\left[1 -\frac{c}{c_\lambda}\right]\, d\xbf\\
\leq & \,\,1-\frac{c}{c_\lambda}.
\end{split}
\end{equation*}
So $\|\pi^\lambda - \pi\|_{\text{TV}} \leq 2(1-\frac{c}{c_\lambda})$. By \citep{rockafellar2009variational}, when $g(\xbf)$ is proper, lower-semicontinuous, and prox-bounded with threshold $\lambda_g >0$, $g^\lambda(\xbf)$ converges pointwise to $g(\xbf)$ as $\lambda \downarrow 0$. Moreover, since $g^{\lambda}(\xbf)$ is pointwise non-decreasing as $\lambda$ decreases, by the monotone convergence theorem, $\lim_{\lambda \downarrow 0} c_\lambda = c$. Thus 
$$
\lim_{\lambda \downarrow 0} \|\pi^\lambda - \pi\|_{\text{TV}} \leq \lim_{\lambda \downarrow 0}  2\left(1-\frac{c}{c_\lambda}\right) = 0.
$$
\end{proof}

Proof of the convergence of MAP follows that in \cite{presman2022distance} and holds under more general conditions than the convergence in total-variation. The convergence in total-variation assumes that $\pi(\thetabf\mid Y)$ defines a proper, nondegenerate density. This is not true when $\pi(\thetabf\mid Y)$ concentrates on a subset of $\mathbb{R}^d$ with Lebesgue measure 0. More work is needed to study the theoretical properties for problems that involve priors with varying dimensionality \citep{XuZhouHuDuan21ProxPrior}.

\section{Details of the Empirical Bayesian Approach}
In this section describe an empirical Bayesian method called  stochastic approximation proximal gradient (SAPG) introduced in \cite{vidal2020maximum}, which calibrates the unknown variance $\sigma^2$ and regularization parameters $\alpha$ by maximum marginal likelihood estimation. \cite{de2020maximum} provides theoretical guarantees for this approach when the variance parameter $\sigma^2$ is known. We adapt Algorithm 4 in \cite{vidal2020maximum}, where both the variance parameter and the regularization parameter are unknown, to the context of matrix completion. More specifically, we use SAPG to estimate the appropriate $\sigma^2$ and $\alpha$ in the following model:
\begin{eqnarray}\label{eq:SAPGmodel}
    \pi(\Xbf|\Ybf) & \propto &\exp\left\{-\frac{\lVert P_{\Omega}(\Ybf-\Xbf) \rVert_F^2}{2\sigma^2}-\alpha \lVert \Xbf\rVert_*\right\}.
\end{eqnarray}
To enable posterior sampling from the nonsmooth density \Eqn{SAPGmodel}, one can replace $\alpha \lVert \Xbf\rVert_*$ with its Moreau envelope and arrive at the following surrogate density:
\begin{eqnarray}\label{eq:SAPGmodelsur}
    \pi^\lambda(\Xbf|\Ybf) & \propto &\exp\left\{-\frac{\lVert P_{\Omega}(\Ybf-\Xbf) \rVert_F^2}{2\sigma^2}-g^\lambda( \Xbf)\right\},
\end{eqnarray}
where 
$$
g^\lambda( \Xbf) = \underset{\Zbf\in \Real^{n\times m}}\min\; \alpha \lVert \Zbf\rVert_*+\frac{1}{2\lambda}\lVert \Zbf - \Xbf\rVert_F^2.
$$
Now that \Eqn{SAPGmodelsur} is smooth, sampling from \Eqn{SAPGmodelsur} can be achieved with the the following Langevin dynamics:
\begin{eqnarray}\label{eq:MYULAtransition}
\Xbf_{l+1} &= &\left(1-\frac{\gamma}{\lambda}\right)\Xbf_{l}-\frac{\gamma}{\sigma^2}P_{\Omega}(\Xbf_{l}-\Ybf) + \frac{\gamma}{\lambda}\operatorname{prox}_g^\lambda(\Xbf_l)+\sqrt{2\gamma}\bm{\zeta}_{l+1},
\end{eqnarray}
where $\bm{\zeta}_{l+1}$ is a $n\times m$ matrix with random normal entries. Simply repeating \Eqn{MYULAtransition} gives us the MYULA algorithm \citep{durmus2018efficient}. We denote the sampling step \Eqn{MYULAtransition} as $\bf{R}_{\gamma,\lambda, \alpha, \sigma^2}(\Xbf_l,\cdot)$. With this transition kernel defined,
we now adapt Algorithm 4 in \cite{vidal2020maximum} to our notation in \Alg{mcsapg}.

\begin{algorithm}[htbp]
\caption{MC-SAPG}\label{alg:mcsapg}
\begin{algorithmic}[1]
\REQUIRE $\alpha_1$, $\sigma^2_1$, $\Xbf_1$, $\delta_l$, $\delta_l^\prime$, $\omega_l$, $\alpha_{\min}$, $\alpha_{\max}$, $\sigma^2_{\min}$, $\sigma^2_{\max}$, $\gamma$, $\lambda$, $N$
\STATE Data: $\Ybf\in\Real^{n\times m}, \Omega=\{(i, j): y_{ij} \text{ is observed}\}$
\FOR{$l=1$ to $N-1$}
  \STATE Sample $\Xbf_{l+1}\sim \bf{R}_{\gamma,\lambda, \alpha_l, \sigma^2_l}(\Xbf_l,\cdot)$
  \STATE Set $\alpha_{l+1} = P_{[\alpha_{\min},\alpha_{\max}]}(\alpha_l+\delta_{l+1}(nm/\alpha_l-\lVert \Xbf_l\rVert_*))$ 
  \STATE Set $\sigma^2_{l+1} = P_{[\sigma^2_{\min},\sigma^2_{\max}]}(\sigma^2_l+\delta_{l+1}^\prime ( \lVert P_{\Omega}(\Ybf-\Xbf_l) \rVert_F^2/(2(\sigma^2_l)^2)-|\Omega|/(2\sigma^2_l)))$
\ENDFOR
\STATE Output: $\bar{\alpha}_N = (\sum_{l=1}^N \omega_l \alpha_l) /(\sum_{l=1}^N \omega_l) $, $\bar{\sigma}^2_N = (\sum_{l=1}^N \omega_l \sigma^2_l) /(\sum_{l=1}^N \omega_l) $
\end{algorithmic}
\end{algorithm}

We reuse the data generation protocol in section 5 to conduct the numerical experiments for SAPG. Following the recommendations in \cite{vidal2020maximum}, we set the parameters in \Alg{mcsapg} as following: $\Xbf_1=P_\Omega(\Ybf)$, $N=1000$, $w_l = 0$ for $l$ from 1 to 500, $w_l = 1$ for $n$ from 501 to 1000, $\delta_l=\delta_l^\prime = 10 (n+1)^{-0.8}/(nm)$, $\alpha_{\min}=1$,  $\alpha_{\max}=10$, $\sigma^2_{\min}=0.1$, $\sigma^2_{\max}=1$, $\lambda = \sigma^2_{\min}$, $\gamma = 0.98(1/\sigma^2_{\min}+1/\lambda)^{-1}$, $\sigma^2_1 = \sigma^2_{\min}$, $\alpha_1=\alpha_{\min}$. After we have obtained $\bar{\alpha}_N$ and $\bar{\sigma}^2_N$, we use the sampling algorithm SK-ROCK \citep{pereyra2020accelerating} to sample from \Eqn{SAPGmodelsur}, where $\alpha=\bar{\alpha}_N$ and $\sigma^2=\bar{\sigma}^2_N$. We direct readers to \cite{pereyra2020accelerating} for the algorithmic details of SK-ROCK. We use the following set of parameters for SK-ROCK: $\lambda=0.1\bar{\sigma}^2_N$, $s=10$ (number of stages) and $\delta=0.8$ (ratio to maximum step size). We perform 2000 SK-ROCK sampling steps, and discard the first 1000 samples as burn-in.  The result is presented in Table 1 of the main text.

\section{Group lasso}

In many applications, predictors form natural groups and sparsity is sought at the group level. Canonical examples are factorial analysis \citep{YuanLin06GroupLasso}, gene association mapping \citep{ZhouLange10LassoGrpMix}, multi-task learning \citep{BachJenattonMairal11SIP}, and multi-response regression. Regularization is achieved by the group lasso penalty $\sum_g w_g \|\betabf_g\|_2$ \citep{YuanLin06GroupLasso}, where $w_g$ are known group weights.

Consider linear regression with the group lasso penalty. Assume $\ybf \in \mathbb{R}^n$ and $\betabf \in \mathbb{R}^p$. The regularized loglikelihood is 
$$
\, - \frac{n}{2} \log \sigma^2 - \frac{\|\ybf-X\betabf\|_2^2}{2\sigma^2} - \rho \sum_g w_g \|\betabf_g\|_2.
$$
Let $\mathcal{E} = \{(\betabf_g, \alpha): \sum_g w_g \|\betabf_g\|_2 \le \alpha, \alpha > 0\}$
and $g(\betabf,\alpha) = \delta_{\mathcal{E}}(\betabf,\alpha)$. With an $IG(r_{\sigma^2},s_{\sigma^2})$ prior for $\sigma^2$ and an $IG(r_\alpha,s_\alpha)$ prior for $\alpha$, the smoothed posterior log-density up to an irrelevant additive constant is 
\begin{equation*}
\begin{split}
& \log \pi^\lambda(\betabf, \log\sigma^2, \log\alpha)\\
= & -(\frac{n}{2} + s_{\sigma^2}) \log \sigma^2 - \frac{\|\ybf-X\betabf\|_2^2 + 2 r_{\sigma^2}}{2\sigma^2}\\
& -s_\alpha \log \alpha - \frac{r_\alpha}{\alpha} - g^\lambda(\betabf, \alpha).
\end{split}
\end{equation*}
The proximal mapping for the group lasso penalty $g(\xbf)=\sum_g w_g \|\xbf_g\|_2$, also called $L_{1,2}$ or $L_{2,1}$ norm, is well known
$$
\prox_g^\lambda(\xbf) = \left( 1 - \frac{\lambda}{w_g \|\xbf\|_g} \right)_+ \xbf_g.
$$
The gradients are
\begin{eqnarray*}
\frac{\partial f}{\partial \betabf} &=& \sigma^{-2} X^T(\ybf - X \betabf) - \lambda^{-1} [\betabf - \prox_g^\lambda(\betabf,\alpha)_{\betabf}], \\
\frac{\partial f}{\partial \log \sigma^2} &=& - (\frac{n}{2} + s_{\sigma^2}) + \frac{\|\ybf-X\betabf\|_2^2 + 2 r_{\sigma^2}}{2\sigma^2}, \\
\frac{\partial f}{\partial \log \alpha} &=& - s_\alpha + \frac{r_\alpha}{\alpha} - \lambda^{-1} \alpha [\alpha - \text{prox}_g^\lambda(\beta,\alpha)_{\alpha}].
\end{eqnarray*}

\begin{figure}[!tbp]
\centering
\includegraphics[width=0.8\textwidth]{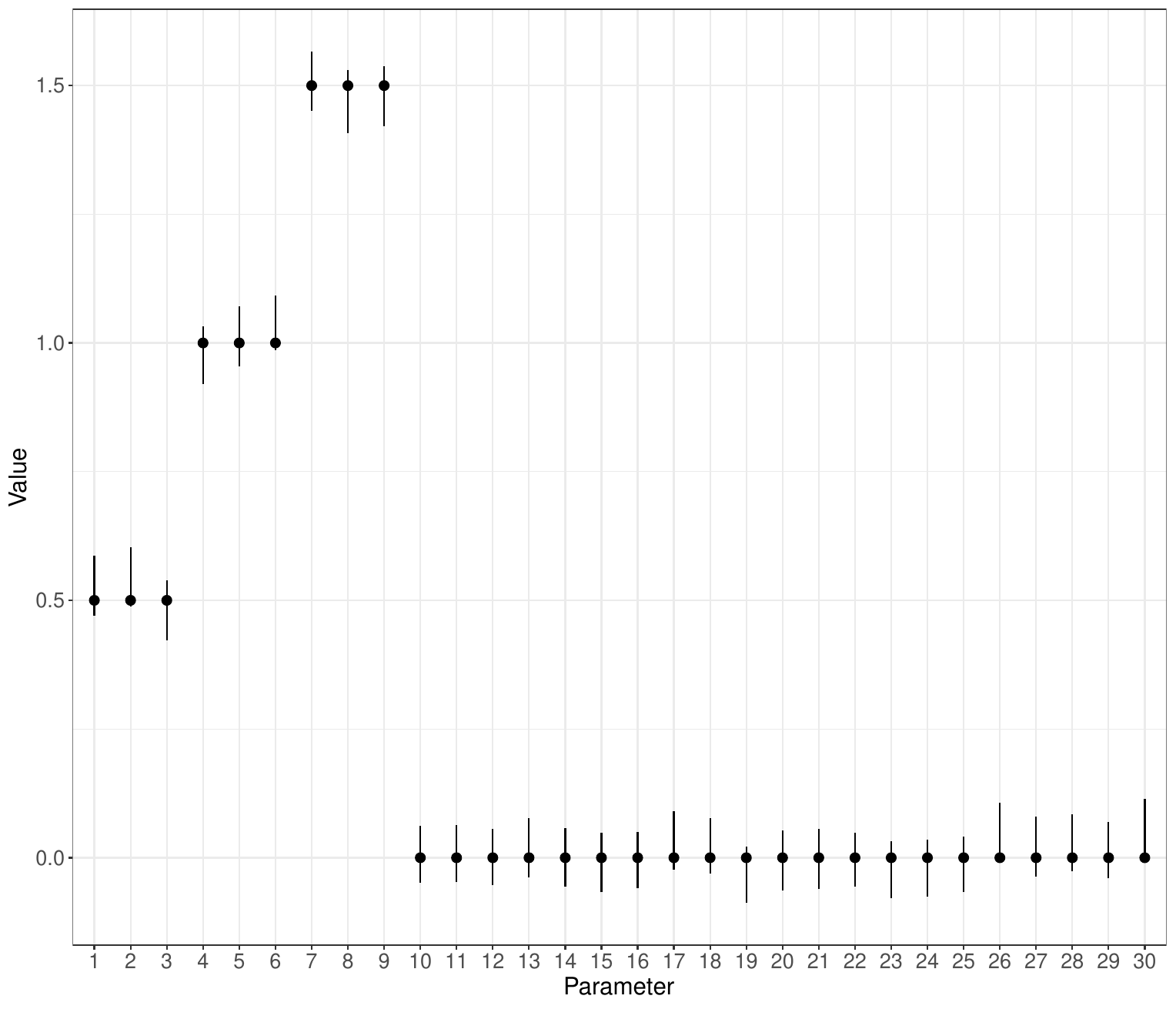}
  \caption{The 95\% credible intervals of ProxMCMC group lasso on the simulated data. }
  \label{fig:group-lasso}
\end{figure}

We illustrate ProxMCMC on group lasso using a simulated data set. Specifically, we generate the response $\ybf$ from $N(\Xbf\betabf, \sigma^2\Ibf)$, where $\Xbf$ is a $300\times30$ matrix generated by i.i.d standard normal, and they form 10 groups with 3 covariates each. The regression coefficient is 
$$\betabf = [0.5, 0.5, 0.5, 1,1,1,1.5,1.5,1.5, 0, ..., 0],$$ and the error standard deviation is $\sigma = 0.5$. We used $IG(0.1, 0.1)$ as a prior for $\sigma^2$ and $IG(1, 30+1)$ as a prior for $\alpha$, set $\lambda = 10^{-6}$, and ran HMC for 10,000 iterations. Figure \ref{fig:group-lasso} shows the 95\% credible intervals for individual parameters, which show excellent coverage of the truth. 

\clearpage
\bibliographystyle{apalike}
\bibliography{proxmcmc.bib}    
\end{document}